\documentclass[sigconf,screen]{acmart}
\AtBeginDocument{%
  }

\newcommand{\qj}{\textcolor{black}}
\newcommand{\ab}{\textcolor{black}}
\usepackage[normalem]{ulem}

\setcopyright{acmlicensed}
\copyrightyear{2018}
\acmYear{2018}
\acmDOI{XXXXXXX.XXXXXXX}
\acmConference[Conference acronym 'XX]{Make sure to enter the correct
  conference title from your rights confirmation email}{June 03--05,
  2018}{Woodstock, NY}
\acmISBN{978-1-4503-XXXX-X/2018/06}

 \usepackage{booktabs}
\usepackage{hyperref}

 \usepackage{multirow}

\usepackage{siunitx} 
\copyrightyear{2026}
\acmYear{2026}
\setcopyright{cc}
\setcctype{by}
\acmConference[CHI '26]{Proceedings of the 2026 CHI Conference on Human Factors in Computing Systems}{April 13--17, 2026}{Barcelona, Spain}
\acmBooktitle{Proceedings of the 2026 CHI Conference on Human Factors in Computing Systems (CHI '26), April 13--17, 2026, Barcelona, Spain}
\acmPrice{}
\acmDOI{10.1145/3772318.3790581}
\acmISBN{979-8-4007-2278-3/2026/04}
\begin{document}

\title[Usage Patterns Impact Presence]{Usage Matters: The Role of Frequency, Duration, and Experience in Presence Formation in Social Virtual Reality}

\author{Qijia Chen}
\orcid{0000-0003-1038-5461}
\affiliation{%
\department{Department of Computer Science}
  \institution{University of Helsinki}
  \city{Helsinki}
   \country{Finland}
}
\email{qijia.chen@helsinki.fi}

\author{Andrea Bellucci}
\orcid{0000-0003-4035-5271}
\affiliation{%
\department{Department of Computer Science and Engineering}
  \institution{Universidad Carlos III de Madrid}
  \city{Leganés}
  \country{Spain}
  }
\email{abellucc@inf.uc3m.es}

\author{Giulio Jacucci}
\orcid{0000-0002-9185-7928}
\affiliation{
\department{Department of Computer Science}
  \institution{University of Helsinki}
  \city{Helsinki}
  \country{Finland}
  }
\email{giulio.jacucci@helsinki.fi }

\renewcommand{\shortauthors}{Qijia et al.}



\begin{abstract}

The sense of presence is central to immersive experiences in Virtual Reality (VR), and particularly salient in socially rich platforms like social VR. While prior studies have explored various aspects related to presence, less is known about how ongoing usage behaviors shape presence in everyday engagement. To address this gap, we examine whether usage intensity, captured through frequency of use, session duration, and years of VR experience, predicts presence in social VR. A survey of 295 users assessed overall, social, spatial, and self-presence using validated scales. Results show that both frequency and duration consistently predict higher presence across all dimensions, with interaction effects indicating that frequent and extended sessions synergistically amplify the experience of “being there.” These effects were stable across age and gender. Our findings extend presence research beyond the laboratory by identifying behavioral predictors in social VR and offer insights for building inclusive environments that reliably foster presence.

\end{abstract}



\maketitle

\section{Introduction}
Presence, commonly described as the psychological state of “being there” in a mediated environment~\cite{steuer1992defining,slater2003note}, has long been considered central to the quality of Virtual Reality (VR) experience. Presence enhances user satisfaction and emotional engagement \cite{Jo2024A, Tian2021}, also supports meaningful outcomes in learning, therapy, and social connection \citep{riva2014interacting, nam2024does}. Prevailing theoretical models conceptualize presence as a multidimensional construct encompassing spatial, social, and self-related dimensions \citep{heater1992being, makransky2017development, lee2004presence}. Each corresponds to different experiential aspects of immersion. Spatial presence reflects the feeling of being located within the virtual environment; social presence captures the sensation of being with others; and self-presence denotes the identification with one’s virtual representation \citep{ratan2013self, biocca2003toward}.

A long line of research shows that presence is strengthened by technical features such as immersive displays, accurate tracking, and haptic feedback \citep{gonccalves2022evaluation, gibbs2022comparison}. More recent studies have started to look at user behavior, for example, how often people use VR (frequency), how long they stay in a session (duration), and how experienced they are (years of use) may shape presence formation. 
The results are mixed. Some studies suggest that repeated or extended exposure enhances immersion through cognitive adaptation \citep{thorp2022temporal, diemer2015impact}, while others find no reliable effect of session duration or repeated use \citep{melo2016impact, porter2022lingering}. 
Nevertheless, most of this research has been conducted in tightly controlled laboratory settings, where behavior is restricted, exposure is brief, and interactions are externally scripted. Such settings may be valuable for isolating causal effects, but provide only a narrow view of presence. They tell us little about how presence develops when people engage voluntarily, repeatedly, and for varying lengths of time in their everyday lives.

Social VR platforms such as VRChat\footnote{\qj{VRChat: \url{https://hello.vrchat.com/}
 (accessed 13 November 2025).}} or Rec Room\footnote{\qj{Rec Room: \url{https://recroom.com/}
 (accessed 10 November 2025)}} offer a way to move past these ecological limitations. Unlike traditional lab-based VR testbeds, they let people engage on their own terms: staying for as long as they like, returning night after night, and interacting through embodied avatars in shared spaces \citep{maloney2021social, sykownik2023vr}. Over time, these environments support communities that develop habits stretching over months or even years. All three dimensions of presence (spatial, social, and self-) are naturally activated through embodied interaction, spatial audio, and co-regulated behavior. Importantly, usage patterns vary widely across users. Some spend hours each day, others drop in occasionally, and their accumulated experience differs dramatically. These natural variations are difficult to reproduce under controlled conditions. 
Early evidence also links presence and embodiment in social VR to outcomes such as enjoyment, self-expansion, and even compulsive use \citep{van2023feelings, barreda2022hooked, barreda2022psychological}, underscoring the value of studying presence where users’ behavior is self-directed and temporally unconstrained.

Building on these perspectives, the present study examines how \emph{usage intensity} in social VR, operationalized as usage frequency, session duration, and years of VR experience, relates to overall presence and to its spatial, social, and self- dimensions. We additionally test whether these relationships vary by age or gender. We ask: 
\begin{itemize}
    \item \textbf{RQ1:} How \qj{do} usage frequency, session duration, and years of VR experience predict overall presence and spatial, social, and self-presence?
   \item \textbf{RQ2:} Do age and gender moderate these associations? 
\end{itemize} 

To address these questions, we surveyed 295 active social VR users about their demographics, usage patterns, and presence experiences using validated, adapted multi-item scales. We then applied hierarchical multiple regression to estimate main effects of usage variables, probe interactions among them (e.g., frequency~$\times$~duration), and assess moderation by age and gender. 

This study advances presence research by moving beyond short-term, scripted lab tasks to capture the everyday patterns of active social VR communities. Our contributions are threefold:
(1) We provide large-scale, survey-based evidence of behavioral predictors of presence across three core dimensions, showing robust effects of frequency and duration as well as interaction effects amplified by prolonged experience; (2) We show that these behavior–presence relationships remain consistent across gender and age, with no evidence of demographic moderation; (3) We provide evidence based insights for creating immersive and inclusive virtual environments that account for behavioral intensity.


\section{Related Works}
This section reviews prior work in \ab{four} areas: theories of presence in VR, studies on how usage behaviors shape presence, \ab{objective measures of presence}, and research positioning social VR as a naturalistic setting. 

\subsection{Dimensions of Presence in Virtual Reality}
The sense of presence in VR refers to the subjective feeling of ``being there'' in a mediated environment \citep{slater2003note,slater1997framework}. Presence is widely regarded as a core construct in immersive media and it is considered critical in applications ranging from gaming and digital communication to education and therapy \citep{riva2014interacting, park2009understanding, lehman2010creating}. Although definitions vary, most accounts describe presence as a sensory-driven, automatically-triggered experience that becomes accessible through introspection \citep{lee2004presence, wirth2007process, barreda2022psychological}. A strong sense of presence enhances user satisfaction, enjoyment \citep{nam2024does}, and task performance \citep{youngblut2003relationship, stevens2015relationship}. Presence also supports emotional and cognitive involvement \cite{Tian2021, kuhne2023}, which is particularly relevant for learning, rehabilitation, and socially meaningful interaction \citep{tang2022experiencing, vesisenaho2019virtual, takac2023cognitive}.

Scholars frame presence as a multidimensional construct~\cite{heater1992being}, typically encompassing three interrelated components: spatial, social, and self-presence~\citep{makransky2017development}.
Spatial presence refers to the sensation of being physically “inside” a virtual or mediated environment, rather than merely observing it on a screen \cite{wirth2007process}. 
Social presence is the extent to which individuals perceive others present or co-located in a mediated environment \cite{biocca2003toward, lowenthal2010social}. It reflects the quality of interpersonal interactions and the feeling of being socially connected in digital spaces \cite{skarbez2017survey}.
Self-presence refers to the extent to which individuals perceive their virtual representation (avatar, digital body, or in-game character) as an extension of themselves \cite{ratan2013self, kilteni2012sense}.

\ab{Together, these three dimensions structure how users experience and act within virtual spaces and provide the conceptual basis for examining how usage patterns shape presence in social VR.}

\subsection{Impact of Usage Patterns on Presence}

A substantial body of research has explored how technological features, such as display immersion \citep{lin2002effects, mcmahan2012evaluating}, tracking fidelity \citep{gonccalves2022evaluation, voigt2020influence}, and haptic feedback \citep{gibbs2022comparison, sallnas2000supporting,kim2017study}, shape users’ sense of presence in virtual environments. Attention also turned to user behaviors and usage patterns, such as how frequently people use VR, and how long their VR sessions last.

Few studies suggest a minimum continuous exposure is needed to ``immerse'' the user. 
\ab{For example, evidence suggests that users adapt cognitively over the first minutes of exposure \citep{diemer2015impact}, with presence increasing before stabilizing \citep{thorp2022temporal}.}
However, these effects are not always consistent. \ab{Other work finds no reliable differences across short exposure intervals, even when duration varies meaningfully \citep{melo2016impact, melo2018presence}, and similar null effects have been reported in non-VR media such as games \citep{lachlan2011experiencing}. Some studies also suggest that excessively long sessions may undermine presence due to fatigue or cognitive overload \citep{2007Effects}. } 
Research on the frequency 
\ab{likewise produced mixed findings}. While it is well known that users often become desensitized to certain physical symptoms such as simulator sickness with repeated exposure \cite{kennedy2000duration,howarth2008characteristics,hill2000habituation}, presence itself appears to remain stable 
\ab{across repeated sessions \citep{bailenson2006longitudinal, porter2022lingering}.} 

\ab{Critically, nearly all prior work has been} conducted in tightly controlled laboratory environments, where exposure is short-term, user behavior is highly scripted, and tasks are externally assigned. While such experimental designs are valuable for isolating causal mechanisms, they may fail to reflect how people engage with VR in their everyday lives: voluntarily, repeatedly, and with varying levels of intensity \citep{9417636}. In particular, real-world use is shaped not just by technology, also by personal motivation, social context, and the freedom to enter and exit at will \citep{bailenson2018experience}. These factors are typically absent from laboratory setups. 
As others have noted \citep{diemer2015impact}, the effects of long-term exposure on presence, such as multi-hour sessions or use across weeks, remain underexplored, highlighting the need for research in more ecologically valid settings.
The emergence of social VR platforms offers a timely and important opportunity to address this gap.

\subsection{\ab{Physiological and Behavioral Measures of Presence}}
\qj{Prior work has also questioned how presence should be measured \citep{slater2004colorful, skarbez2017survey}. Presence is most commonly assessed through self-report questionnaires that ask participants to retrospectively rate how much they ``felt there'' during a VR experience \citep{schubert2001experience}. While such instruments are practical and have been extensively validated, they remain subjective and depend on participants' introspection and truthful self-assessment \citep{ grassini2020questionnaire}. In response, researchers have explored more ``objective'' or indirect indicators of presence. }
\qj{Physiological measures such as heart rate, skin conductance, and other autonomic responses can capture continuous, real-time reactions without interrupting the VR experience \citep{meehan2002physiological}. These indices often covary with changes in immersion or emotional intensity, making them appealing for presence research \citep{grassini2020questionnaire}. However, recent reviews note that physiological responses are sensitive to many processes beyond presence (such as stress, attention, or cognitive load) and therefore require careful interpretation \citep{grassini2020questionnaire}.}
\ab{Behavioral indicators, such as head or body movement patterns, have similarly been proposed as unobtrusive signals that correlate with immersion or engagement \citep{mcmahan2012evaluating, gonccalves2022evaluation}. Yet these measures remain underutilized in everyday VR settings, where technical access to detailed interaction logs is often limited outside controlled laboratory environments \citep{maloney2021social}.}

\qj{Building on this line of work, our study focuses on usage intensity as another relatively objective, behaviorally grounded indicator that, in principle, can be captured via system logs (e.g., frequency and duration of social VR sessions). Although these measures are self-reported in the present survey, they directly reflect observable usage patterns and complement traditional questionnaire-based assessments of presence, especially in large-scale, ecologically valid studies of everyday VR use.
Situating usage-based measures alongside physiological and behavioral indicators helps broaden the methodological toolkit for studying presence in social VR.}

\subsection{Social VR as a Research Setting}
Social VR platforms offer an immersive environment where users can interact with one another through embodied avatars \citep{Maloney2020, maloney2021social}. 
\ab{Through features such as spatial audio, real-time motion tracking, and shared interactive spaces, these platforms enable rich interpersonal experiences that naturally activate spatial, social, and self-presence \citep{bailenson2018experience, tanenbaum2020make, wienrich2018social}.}

The increasing accessibility of consumer-grade VR hardware (e.g., Meta Quest, HTC Vive) has fueled the rapid growth of social VR platforms such as VRChat or Rec Room. 
These environments support a diverse range of user-driven activities, from casual meetups and role-playing to dance parties, live performances, and collaborative games \citep{freeman2021body, chen2024d, piitulainen2022vibing}. Importantly, such activities often transcend entertainment. They involve personal expression, emotional resonance, and mutual engagement, providing rich opportunities for users to fulfill different needs \citep{Maloney2020, Hugging5, 9417636}.

Social VR has \ab{also become an important setting for studying social behavior and immersive psychological processes. Because behavior unfolds voluntarily rather than through scripted tasks \citep{maloney2021social}, researchers have used social VR to examine naturally occurring phenomena such as phantom touch \cite{chen2024understanding}, mirror dwelling \citep{fu2023mirror,chen2025mirror}, and other emergent nonverbal communication norms \citep{Maloney2020,chen2025Mutes}.} 
In addition, social VR \ab{is} an important ground for studying the sense of presence. \ab{Its ecological, self-directed nature enables presence to be examined as a naturally occurring experience rather than one induced through short-term laboratory tasks.}
A growing body of empirical work underscores \ab{this potential}. 
For instance, recent studies in social VR show that spatial and social presence predict meaningful psychological outcomes such as enjoyment and self-expansion \citep{van2023feelings}, as well as social well-being \citep{barreda2022psychological}.

Crucially, social VR provides a unique opportunity to observe the three dimensions of presence interacting in realistic use. 
Unlike laboratory studies (often limited by small samples, brief exposure times, and scripted tasks), social VR involves large and diverse user groups who engage on their own terms. Users may drop in briefly, stay for several hours, or return habitually across weeks or months. This behavioral diversity provides a natural context in which presence can be examined as a natural state shaped by actual usage.
By drawing on survey data from active social VR users, we move beyond the constraints of lab experiments and provide ecologically grounded insights into how immersive experiences are sustained through everyday engagement.

\section{Method}
This study employed an online survey targeting individuals with prior experience in social VR. The final sample ($N = 295$) provided information on demographics, VR usage behaviors, and subjective sense of presence, measured using adapted, validated scales. Reliability and validity of the presence measures were confirmed through psychometric analyses. To assess the influence of demographic and behavioral factors on presence, we conducted hierarchical multiple regressions, including interaction and moderation effects. 

\subsection{Data Collection}
We recruited participants through Prolific, a crowdsourcing platform widely used in academic research. Eligibility required respondents to be at least 18 years old and proficient in English. 
To ensure familiarity with the context, participants were given a brief description of social VR (adapted from \cite{schulenberg2024does}) that outlined its core features, including avatars, head-mounted displays, and popular platforms such as VRChat and Rec Room. Only those with prior experience in social VR could proceed to the main survey.
A total of 341 responses were collected. After removing incomplete submissions and low-quality cases (e.g., straight-lining), the final dataset comprised 295 valid responses.

\subsection{Measurement}
Demographic information relating to gender, age, and geographic location was collected. Participants were asked to indicate their gender by selecting ‘man,’ ‘woman,’ ‘non-binary,’ or ‘prefer not to say.’ One participant selected ‘non-binary,’ and no participants selected ‘prefer not to say.’ The sample was predominantly male (72.9\%), with a smaller proportion identifying as female (26.8\%). Age was assessed using a drop-down menu, with the majority of respondents falling within the 25–34 age group (47.5\%), followed by 35–44 (20.7\%), and 18–24 (18.3\%). Participants aged 55 or older made up a small portion of the sample (3.0\% combined). Participants were also asked to indicate their geographic region of residence, with responses later recoded into five categories: Europe, Americas, Africa, Asia, and Oceania. The sample was primarily based in Europe (67.8\%), followed by the Americas (18.0\%), Africa (9.8\%), Asia (2.7\%), and Oceania (1.7\%). Participant demographics are summarized in \autoref{tab:demographics} provided in Appendix~\ref{AppendixA}.

Participants were asked to report their VR usage behaviors, including usage frequency, average session duration, and overall experience with social VR systems. Frequency of use was assessed with options ranging from “daily” to “ Not at all.” The majority of participants indicated using VR either a few times per month (44.1\%) or once a month or less (33.9\%), while only a small number reported daily use (1.0\%). Session duration was measured using predefined intervals. Most respondents reported typical session lengths of 30–60 minutes (45.4\%), with a smaller proportion indicating usage durations of 1–2 hours (30.2\%) or more. VR experience was measured in terms of total time spent using VR systems. Over half of the sample had used VR for more than one year: 38.0\% reported 1–3 years of experience, while 21.4\% indicated having more than 3 years of use. A minority of participants were new to VR, with just 3.1\% reporting less than one month of experience. These usage patterns are summarized in \autoref{tab:vr_usage_patterns}, which is presented in Appendix~\ref{AppendixB}.

Social presence was measured with the five-item subscale adapted from the Multimodal Presence Scale for VR environments \citep{makransky2017development}. An example item is: “When I use social VR, I feel like I am in the presence of another person in the virtual environment.”
Spatial presence was assessed using the four-item self-location subscale of the Spatial Presence Experience Scale \citep{hartmann2015spatial}. Items were adapted to reflect experiences in social VR, such as “When I use social VR, it feels as though I am physically present in the virtual environment.”
Self-presence was captured with the five-item self-presence subscale of the Multimodal Presence Scale \citep{makransky2017development}. For example: “When I use social VR, I feel like my avatar is an extension of my real body within the virtual environment.”
\qj{These instruments were chosen because, together, they capture the three dimensions of presence that frame this study: social, self-, and spatial presence. Many other widely used presence questionnaires in VR research (e.g., Presence Questionnaire (PQ) \citep{witmer1998measuring}, Igroup Presence Questionnaire (IPQ) \citep{schubert2001experience}, Slater–Usoh–Steed (SUS) \citep{usoh2000using}) primarily focus on spatial or overall presence and therefore do not concurrently cover all three dimensions in a form suitable for heterogeneous, everyday social VR experiences. Our choice also aligns with recent large-scale survey work on social VR that used the same presence components to predict psychological outcomes and reported good reliability and construct validity \citep{barreda2022psychological,van2023feelings,chen2026presence}.}

All presence scales were rated on a 5-point Likert scale ranging from 1 (strongly disagree) to 5 (strongly agree).
\qj{Minor adaptations were made to fit our study context: items were reworded to refer explicitly to “social VR”. We did not change the underlying item content or the assignment of items to constructs. 
To ensure the suitability of these measures for the regression analyses, we assessed internal consistency and convergent validity.} 
As reported in \autoref{tab:presence-measurements} (Appendix~\ref{AppendixC}), all Cronbach’s alpha and composite reliability (CR) values exceeded the commonly recommended threshold of .70 \citep{nunnally1994psychometric}, indicating good reliability. Convergent validity was evaluated using average variance extracted (AVE), with all AVE values above the .50 cutoff \citep{hair2009multivariate}, suggesting that the constructs capture sufficient variance from their respective items.
Furthermore, all standardized factor loadings were above .70, indicating that each item loaded strongly onto its respective construct \citep{chin2003partial}. Taken together, these results support the reliability and construct validity of the presence measures used in the hierarchical regression analyses. In addition, an overall presence score was created by averaging the three subdimension scores (social, spatial, self-presence), consistent with prior research treating presence as a higher-order construct.

\subsection{Data Analysis}

To examine the impacts of usage patterns on users’ sense of presence, we employed \textit{hierarchical multiple regression analysis}. This method enables the sequential entry of predictor variables into the model in theoretically justified blocks, thereby allowing for the isolation of each block's contribution to variance explained in the dependent variable \citep{cohen2013applied}. Compared to standard multiple regression, this approach offers greater control over covariates and allows researchers to observe changes in $R^2$ as new variables are added.

Before conducting hierarchical regressions, we examined bivariate correlations among key predictors (i.e., frequency, duration, years of use, age, and gender). All correlations were within acceptable ranges, with no indications of multicollinearity (see \autoref{tab:co}). This justified the inclusion of these variables in the same regression models.

Next, we conducted hierarchical multiple regression analyses to examine the predictors of overall sense of presence as well as its subdimensions: social presence, spatial presence, and self-presence. All statistical analyses were performed in R. We utilized several specialized packages to support regression modeling and visualization: \texttt{psych} for reliability analysis, \texttt{lm.beta} for obtaining standardized beta coefficients, \texttt{stargazer} for producing publication-quality regression tables, and \texttt{interactions} for probing and visualizing moderation effects.

Each regression model included four steps. Gender (dummy-coded: 1 = female, 0 = male) and age (z-standardized) were entered as control variables in Step 1. In Step 2, three behavioral predictors—usage frequency, duration per session, and total length of usage (all z-standardized)—were added. In Step 3, we added two-way interaction terms between behavioral predictors (e.g., frequency × duration, frequency × years).
Finally, Step 4 tested moderation by gender and age through additional interaction terms (e.g., gender × frequency, age × duration).
Model performance was assessed using adjusted $R^2$ and $\Delta R^2$ at each step to evaluate the incremental explanatory power. Significance thresholds were set at $p < .05$, with $p < .10$ reported as marginally significant. All predictors were mean-centered prior to interaction term computation to reduce multicollinearity.

\begin{table}[htbp]
\centering
\renewcommand{\arraystretch}{1.2}
\caption{\textbf{Bivariate Correlations Between Independent and Control Variables}}
\label{tab:co}
\begin{tabular}{l
                S[table-format=1.3]
                S[table-format=1.3]
                S[table-format=1.3]
                S[table-format=1.3]
                S[table-format=1.3]}
\toprule
\textbf{Variable} & \textbf{Gender} & \textbf{Frequency} & \textbf{Duration} & \textbf{Years} & \textbf{Age} \\
\midrule
Gender     & 1.000 &  0.044   & -0.142$^{\dagger}$ & -0.235$^{***}$ &  0.061 \\
Frequency  &       &  1.000   &  0.316$^{***}$     &  0.168$^{*}$   &  0.042 \\
Duration   &       &          &  1.000             &  0.264$^{***}$ & -0.122$^{*}$ \\
Years      &       &          &                    &  1.000         &  0.070 \\
Age        &       &          &                    &                &  1.000 \\
\bottomrule
\end{tabular}

\vspace{2mm}
\footnotesize
\textit{Note.} \qj{$\dagger p<.10$}, * $p<.05$, ** $p<.01$, *** $p<.001$.  
\end{table}

\section{Results}
This section presents the results of the analyses examining how usage behaviors predict users’ sense of presence in social VR. We report findings separately for overall presence and its three subdimensions, showing main effects, interaction terms, and moderation by gender and age.

\begin{table*}[t]
\centering
\renewcommand{\arraystretch}{1.2}
\caption{\textbf{Hierarchical Regression Predicting Sense of Presence}}
\label{tab:sense_presence_regression}

\resizebox{.7\textwidth}{!}{%
\begin{tabular}{lcccccccc}
\toprule
\textbf{Variable}
& \multicolumn{2}{c}{\textbf{Model 1}}
& \multicolumn{2}{c}{\textbf{Model 2}}
& \multicolumn{2}{c}{\textbf{Model 3}}
& \multicolumn{2}{c}{\textbf{Model 4}} \\
\cmidrule(lr){2-3}\cmidrule(lr){4-5}\cmidrule(lr){6-7}\cmidrule(lr){8-9}
 & $B$ & $\beta$ & $B$ & $\beta$ & $B$ & $\beta$ & $B$ & $\beta$ \\
\midrule
Female            & 0.328**  & 0.148   & 0.317**  & 0.143   & 0.322**  & 0.145   & 0.342*** & 0.154   \\
Age (z)           & 0.055    & 0.056   & 0.047    & 0.048   & 0.055    & 0.055   & 0.053    & 0.054   \\
Frequency (z)     & 0.198*** & 0.201   & 0.182*** & 0.185   & 0.195*** & 0.197   & 0.189*** & 0.192   \\
Duration (z)      & 0.181*** & 0.183   & 0.169*** & 0.171   & 0.175*** & 0.177   & 0.172*** & 0.174   \\
Years (z)         & 0.008    & 0.008   & 0.007    & 0.007   & 0.013    & 0.014   & 0.039    & 0.039   \\
Freq $\times$ Dur &          &         & 0.138*** & 0.150   &          &         &          &         \\
Freq $\times$ Yr  &          &         &          &         & 0.098*   & 0.101   &          &         \\
Dur $\times$ Yr   &          &         &          &         &          &         & 0.140*** & 0.146   \\
\midrule
$R^2$             & \multicolumn{2}{c}{0.117}   & \multicolumn{2}{c}{0.139}   & \multicolumn{2}{c}{0.127}   & \multicolumn{2}{c}{0.138} \\
$\Delta R^2$      & \multicolumn{2}{c}{--}      & \multicolumn{2}{c}{0.022**} & \multicolumn{2}{c}{\qj{0.010$^{\dagger}$}}& \multicolumn{2}{c}{0.021**} \\
\bottomrule
\multicolumn{9}{l}{\footnotesize \textit{Note.}  $\dagger p<.10$, * $p < .05$, ** $p < .01$, *** $p < .001$. Standardized $\beta$ coefficients shown without significance markers.}
\end{tabular}}
\end{table*}

\begin{table*}[htbp]
\centering
\small
\setlength{\tabcolsep}{4pt}        
\renewcommand{\arraystretch}{1.1}  
\caption{\textbf{Hierarchical Regression: Gender and Age Interactions Predicting Sense of Presence}}
\label{tab:interaction-regression}


\begin{tabular}{lcccccccccccc}
\toprule
\textbf{Variable} & \multicolumn{2}{c}{\textbf{Model 1}} & \multicolumn{2}{c}{\textbf{Model 2}} & \multicolumn{2}{c}{\textbf{Model 3}} & \multicolumn{2}{c}{\textbf{Model 4}} & \multicolumn{2}{c}{\textbf{Model 5}} & \multicolumn{2}{c}{\textbf{Model 6}} \\
\cmidrule(lr){2-3} \cmidrule(lr){4-5} \cmidrule(lr){6-7} \cmidrule(lr){8-9} \cmidrule(lr){10-11} \cmidrule(lr){12-13}
 & $B$ & $\beta$ & $B$ & $\beta$ & $B$ & $\beta$ & $B$ & $\beta$ & $B$ & $\beta$ & $B$ & $\beta$ \\
\midrule
Female             & 0.335*** & 0.151 & 0.301**  & 0.135 & 0.289**  & 0.130 & 0.314**  & 0.142 & 0.324**  & 0.146 & 0.329**  & 0.148 \\
Age (z)            & 0.052    & 0.052 & 0.051    & 0.051 & 0.052    & 0.053 & 0.058    & 0.059 & 0.055    & 0.056 & 0.056    & 0.057 \\
Frequency (z)      & 0.224*** & 0.227 & 0.203*** & 0.206 & 0.203*** & 0.205 & 0.204*** & 0.206 & 0.200*** & 0.203 & 0.198*** & 0.201 \\
Duration (z)       & 0.181*** & 0.183 & 0.225*** & 0.228 & 0.174*** & 0.177 & 0.183*** & 0.186 & 0.179*** & 0.182 & 0.181*** & 0.184 \\
Years (z)          & 0.011    & 0.011 & 0.006    & 0.006 & 0.063    & 0.064 & 0.005    & 0.005 & 0.008    & 0.008 & 0.008    & 0.008 \\
G $\times$ Freq    & $-$0.099 & $-$0.052 &         &       &         &       &         &       &         &       &         &       \\
G $\times$ Dur     &          &        & $-$0.180 & $-$0.093 &         &       &         &       &         &       &         &       \\
G $\times$ Years   &          &        &          &        & $-$0.175 & $-$0.100 &         &       &         &       &         &       \\
A $\times$ Freq    &          &        &          &        &          &        & $-$0.076 & $-$0.073 &         &       &         &       \\
A $\times$ Dur     &          &        &          &        &          &        &          &        & $-$0.023 & $-$0.023 &         &       \\
A $\times$ Years   &          &        &          &        &          &        &          &        &          &        & 0.014    & 0.013 \\
\midrule
$R^2$              & \multicolumn{2}{c}{0.119} & \multicolumn{2}{c}{0.123} & \multicolumn{2}{c}{0.124} & \multicolumn{2}{c}{0.123} & \multicolumn{2}{c}{0.118} & \multicolumn{2}{c}{0.117} \\
$\Delta R^2$       & \multicolumn{2}{c}{0.002} & \multicolumn{2}{c}{0.006} & \multicolumn{2}{c}{0.007} & \multicolumn{2}{c}{0.006} & \multicolumn{2}{c}{0.001} & \multicolumn{2}{c}{0.000} \\
\bottomrule
\multicolumn{13}{l}{\footnotesize \textit{Note.} $\dagger p<.10$, * $p < .05$, ** $p < .01$, *** $p < .001$. Standardized $\beta$ shown without significance markers.}
\end{tabular}
\end{table*}

\subsection{Models Predicting Overall Sense of Presence}
We conducted hierarchical multiple regression analyses to examine how demographic and behavioral variables predict the overall sense of presence in social VR, followed by interaction analyses involving usage behaviors and demographic moderators.

\subsubsection{Main and Interaction Effects of Usage Behaviors}

\begin{table*}[t]
\centering
\renewcommand{\arraystretch}{1.2}

\caption{\textbf{Hierarchical Regression Predicting Social Presence}}
\label{tab:social_presence_regression}

\resizebox{.7\textwidth}{!}{%
\begin{tabular}{lcccccccc}
\toprule
\textbf{Variable} & \multicolumn{2}{c}{\textbf{Model 1}} & \multicolumn{2}{c}{\textbf{Model 2}} & \multicolumn{2}{c}{\textbf{Model 3}} & \multicolumn{2}{c}{\textbf{Model 4}} \\
\cmidrule(lr){2-3} \cmidrule(lr){4-5} \cmidrule(lr){6-7} \cmidrule(lr){8-9}
 & $B$ & $\beta$ & $B$ & $\beta$ & $B$ & $\beta$ & $B$ & $\beta$ \\
\midrule
Female            & 0.370*** & 0.166 & 0.365*** & 0.164 & 0.367*** & 0.165 & 0.380*** & 0.171 \\
Age (z)           & 0.047    & 0.048 & 0.044    & 0.044 & 0.047    & 0.047 & 0.045    & 0.046 \\
Frequency (z)     & 0.167*** & 0.169 & 0.161*** & 0.162 & 0.165*** & 0.167 & 0.160*** & 0.162 \\
Duration (z)      & 0.207*** & 0.209 & 0.202*** & 0.204 & 0.205*** & 0.207 & 0.200*** & 0.202 \\
Years (z)         & 0.027    & 0.027 & 0.026    & 0.027 & 0.029    & 0.029 & 0.050    & 0.050 \\
Freq $\times$ Dur &          &       & 0.056    & 0.060 &          &       &          &       \\
Freq $\times$ Yr  &          &       &          &       & 0.044    & 0.045 &          &       \\
Dur $\times$ Yr   &          &       &          &       &          &       & 0.107**  & 0.111 \\
\midrule
$R^2$             & \multicolumn{2}{c}{0.119} & \multicolumn{2}{c}{0.123} & \multicolumn{2}{c}{0.121} & \multicolumn{2}{c}{0.131} \\
$\Delta R^2$      & \multicolumn{2}{c}{--}    & \multicolumn{2}{c}{0.004} & \multicolumn{2}{c}{0.002} & \multicolumn{2}{c}{0.012*} \\
\bottomrule
\multicolumn{9}{l}{\footnotesize \textit{Note.}  $\dagger p<.10$, * $p < .05$, ** $p < .01$, *** $p < .001$. Standardized $\beta$ coefficients shown without significance markers.}
\end{tabular}}
\end{table*}

\autoref{tab:sense_presence_regression} summarizes the results of hierarchical regression models predicting users’ overall sense of presence.

Model 1 included gender, age, and the three usage behavior variables. Gender was a significant predictor ($\beta = .148$, $p < .01$), with female users reporting higher presence levels than male users. Frequency ($\beta = .201$, $p < .001$) and duration ($\beta = .183$, $p < .001$) were also significant positive predictors, while age and years of VR experience were not significant. 

Model 2 added the interaction between frequency and duration, which was statistically significant ($\beta = .150$, $p < .001$). This interaction indicates that users who engaged in VR both frequently and for longer sessions reported a stronger sense of presence. 
Model 3 replaced the previous interaction with frequency $\times$ years of experience, which was significant at the $p < .05$ level ($\beta = .101$). This suggests that the positive association between frequency and presence was more pronounced among long-term users. 
Model 4 tested the interaction between duration and years of experience, which also emerged as significant ($\beta = .146$, $p < .001$). This finding implies that long-term users benefit more from extended VR sessions in terms of presence experience. The interaction plots are visualized in \autoref{fig:interaction-effects} (see Appendix~\ref{AppendixD}).

\subsubsection{Moderation Effects of Gender and Age}

\begin{figure}[H]
    \centering
    \includegraphics[width=1\linewidth]{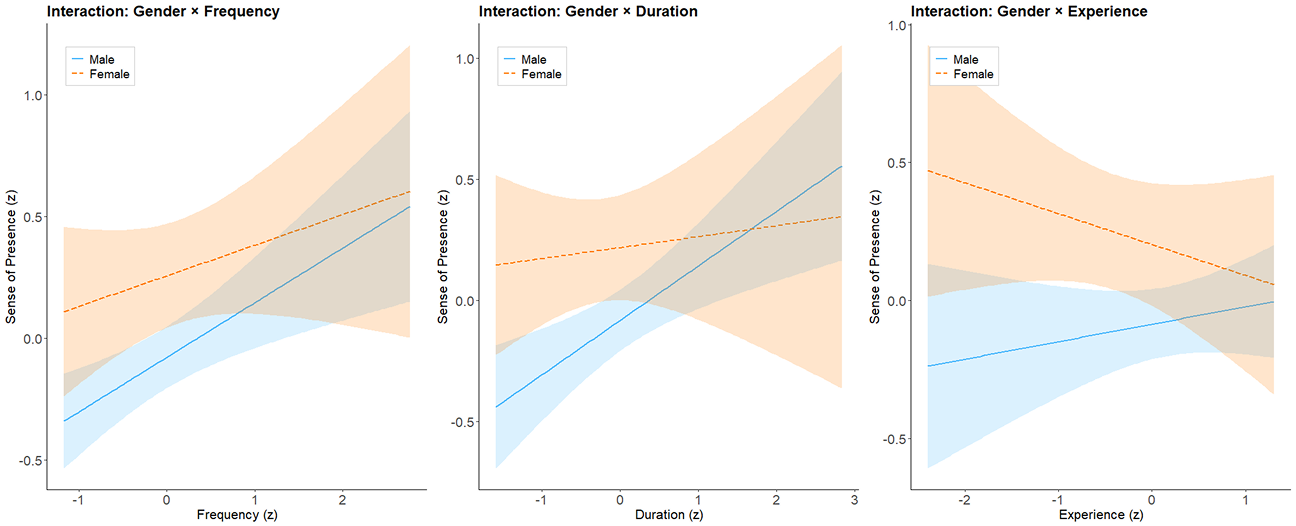}
    \caption{\textbf{Interaction effects between gender and usage behaviors on sense of presence.
    (Left) Gender $\times$ Frequency, (Middle) Gender $\times$ Duration, (Right) Gender $\times$ Experience. 
    Shaded regions represent 95\% confidence intervals.}}
    \label{fig:gender-interaction}
\end{figure}

\begin{figure}[H]
    \centering
    \includegraphics[width=1\linewidth]{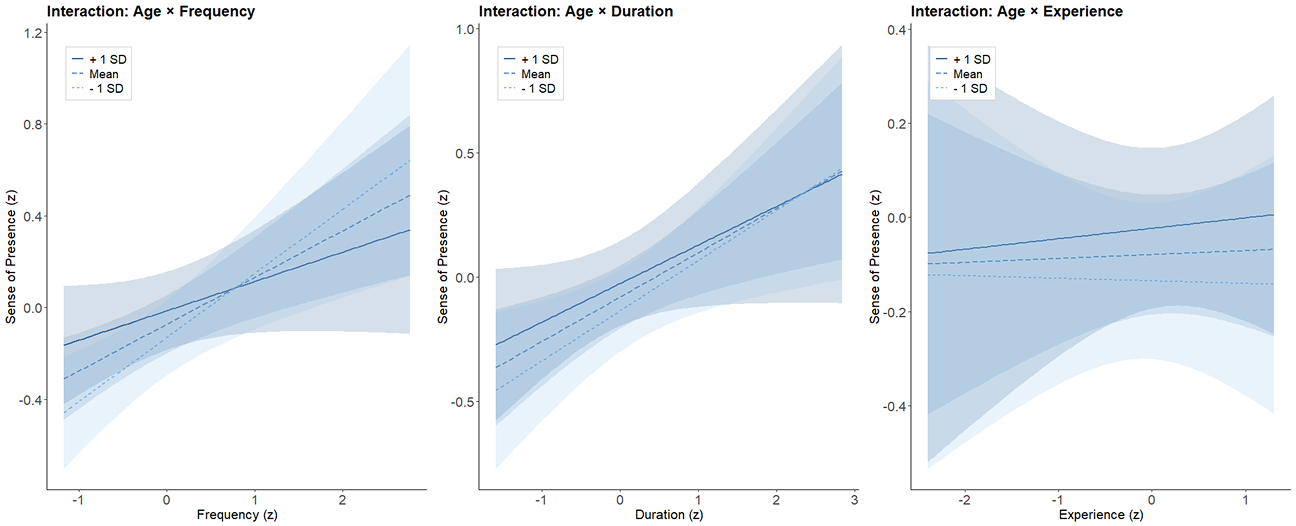}
    \caption{\textbf{Interaction effects between age and usage behaviors on sense of presence.
    (Left) Age $\times$ Frequency, (Middle) Age $\times$ Duration, (Right) Age $\times$ Experience. 
    Shaded regions represent 95\% confidence intervals.}}
    \label{fig:age-interaction}
\end{figure}

\autoref{tab:interaction-regression} summarizes additional regression models examining moderation effects of gender and age.
Across all six models tested, none of the interaction terms reached significance. While a few coefficients (e.g., Gender × Duration) showed marginal trends, the added explanatory power was small. These results suggest that the influence of frequency and duration on presence holds consistently across gender and age subgroups. Additional plots illustrating the moderating effects are provided in \autoref{fig:gender-interaction} and \autoref{fig:age-interaction} (Appendix~\ref{AppendixD}).

\subsection{Models Predicting Social Presence}
This section focuses on predictors of social presence, examining both main effects and interaction patterns. The analytic procedure mirrored that of overall presence.

\subsubsection{Main and Interaction Effects of Usage Behaviors}

\autoref{tab:social_presence_regression} presents the predicting social presence.
Model 1 identified frequency and duration as significant and positive predictors of social presence. Higher usage frequency ($\beta = .169$, $p < .001$) and longer session durations ($\beta = .209$, $p < .001$) were associated with stronger perceptions of social presence. Gender was again significant, with female participants reporting higher levels ($\beta = .166$, $p < .001$), while age and years of VR use did not contribute meaningfully.

\begin{figure}[H]
    \centering
    \includegraphics[width=1\linewidth]{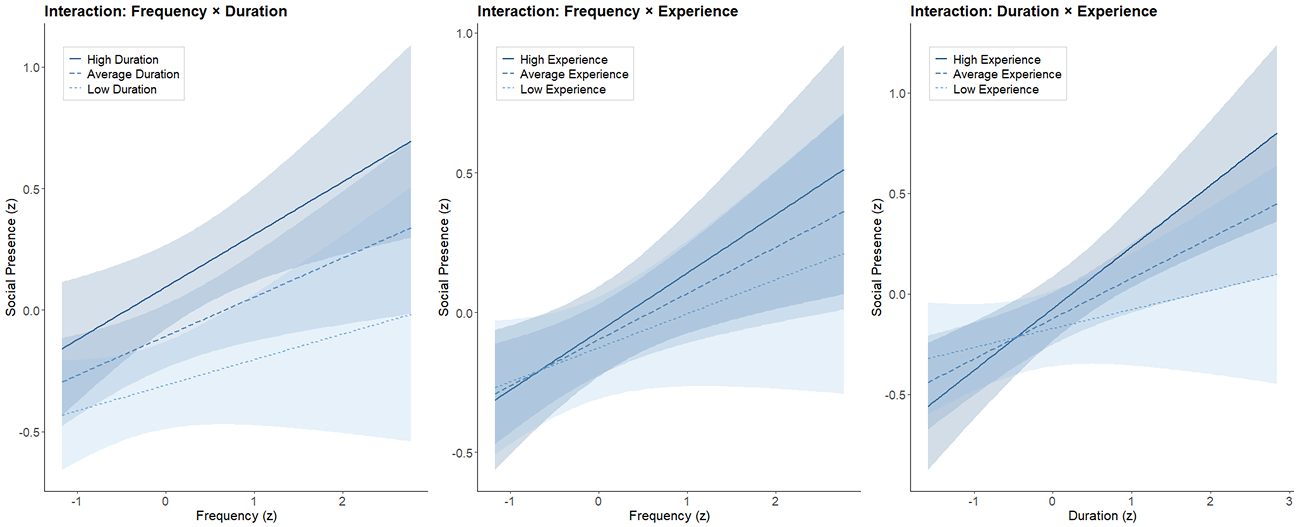}
    \caption{\textbf{Interaction effects of usage behaviors on social presence. 
    (Left) Frequency $\times$ Duration interaction; (Middle) Frequency $\times$ Experience interaction; (Right) Duration $\times$ Experience interaction. 
    All predictors are standardized. Shaded areas represent 95\% confidence intervals.}}
    \label{fig:sp}
\end{figure}

Model 2 included the frequency × duration interaction, but this term did not reach significance. Similarly, Model 3 introduced the frequency × years of experience interaction, which remained nonsignificant. In contrast, Model 4 showed a significant interaction between duration and years ($\beta = .111$, $p < .01$), suggesting that users with more VR experience derive greater social presence from extended sessions.
The interaction analyses are also visualized in \autoref{fig:sp} (Appendix~\ref{AppendixD}).

\begin{table*}[htbp]
\centering
\small
\setlength{\tabcolsep}{4pt}        
\renewcommand{\arraystretch}{1.1}  
\caption{\textbf{Hierarchical Regression: Gender and Age Interactions Predicting Social Presence}}
\label{tab:socialpresence-interaction}
\begin{tabular}{lcccccccccccc}
\toprule
\textbf{Variable} & \multicolumn{2}{c}{\textbf{Model 1}} & \multicolumn{2}{c}{\textbf{Model 2}} & \multicolumn{2}{c}{\textbf{Model 3}} & \multicolumn{2}{c}{\textbf{Model 4}} & \multicolumn{2}{c}{\textbf{Model 5}} & \multicolumn{2}{c}{\textbf{Model 6}} \\
\cmidrule(lr){2-3} \cmidrule(lr){4-5} \cmidrule(lr){6-7} \cmidrule(lr){8-9} \cmidrule(lr){10-11} \cmidrule(lr){12-13}
 & $B$ & $\beta$ & $B$ & $\beta$ & $B$ & $\beta$ & $B$ & $\beta$ & $B$ & $\beta$ & $B$ & $\beta$ \\
\midrule
Female              & 0.372*** & 0.167 & 0.352*** & 0.158 & 0.329**  & 0.148 & 0.352*** & 0.158 & 0.363*** & 0.163 & 0.372*** & 0.167 \\
Age (z)             & 0.046    & 0.046 & 0.044    & 0.045 & 0.044    & 0.045 & 0.051    & 0.052 & 0.046    & 0.047 & 0.049    & 0.049 \\
Frequency (z)       & 0.176*** & 0.178 & 0.170*** & 0.172 & 0.172*** & 0.173 & 0.174*** & 0.176 & 0.171*** & 0.172 & 0.167*** & 0.169 \\
Duration (z)        & 0.208*** & 0.209 & 0.236*** & 0.238 & 0.201*** & 0.203 & 0.211*** & 0.213 & 0.204*** & 0.205 & 0.208*** & 0.209 \\
Years (z)           & 0.028    & 0.028 & 0.025    & 0.026 & 0.084    & 0.085 & 0.022    & 0.022 & 0.026    & 0.026 & 0.027    & 0.027 \\
G $\times$ Freq     & $-$0.035 & $-$0.018 &         &        &         &        &         &        &         &        &         &        \\
G $\times$ Dur      &          &         & $-$0.118 & $-$0.061 &         &        &         &        &         &        &         &        \\
G $\times$ Years    &          &         &          &         & $-$0.182 & $-$0.104 &         &        &         &        &         &        \\
A $\times$ Freq     &          &         &          &         &          &         & $-$0.094 & $-$0.091 &         &        &         &        \\
A $\times$ Dur      &          &         &          &         &          &         &          &         & $-$0.041 & $-$0.041 &         &        \\
A $\times$ Years    &          &         &          &         &          &         &          &         &          &         & 0.027    & 0.026 \\
\midrule
$R^2$               & \multicolumn{2}{c}{0.120} & \multicolumn{2}{c}{0.122} & \multicolumn{2}{c}{0.126} & \multicolumn{2}{c}{0.128} & \multicolumn{2}{c}{0.121} & \multicolumn{2}{c}{0.120} \\
$\Delta R^2$        & \multicolumn{2}{c}{0.001} & \multicolumn{2}{c}{0.002} & \multicolumn{2}{c}{0.006} & \multicolumn{2}{c}{0.009} & \multicolumn{2}{c}{0.002} & \multicolumn{2}{c}{0.001} \\
\bottomrule
\multicolumn{13}{l}{\footnotesize \textit{Note.} \qj{$\dagger p<.10$}, * $p < .05$, ** $p < .01$, *** $p < .001$. Standardized $\beta$ shown without significance markers.}
\end{tabular}
\end{table*}

\subsubsection{Moderation Effects of Gender and Age}
\begin{figure}[H]
    \centering
    \includegraphics[width=1\linewidth]{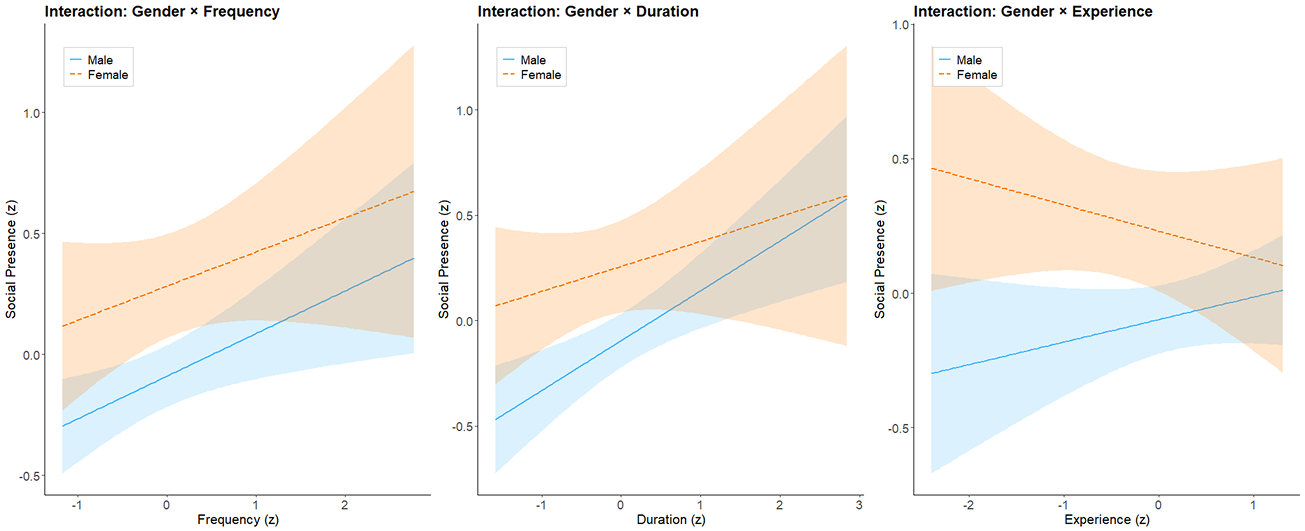}
    \caption{\textbf{Interaction effects between gender and usage behaviors on social presence.
    (Left) Gender $\times$ Frequency, (Middle) Gender $\times$ Duration, (Right) Gender $\times$ Experience. 
    Shaded regions represent 95\% confidence intervals.}}
    \label{fig:gendersocial}
\end{figure}

\begin{figure}[H]
    \centering
    \includegraphics[width=1\linewidth]{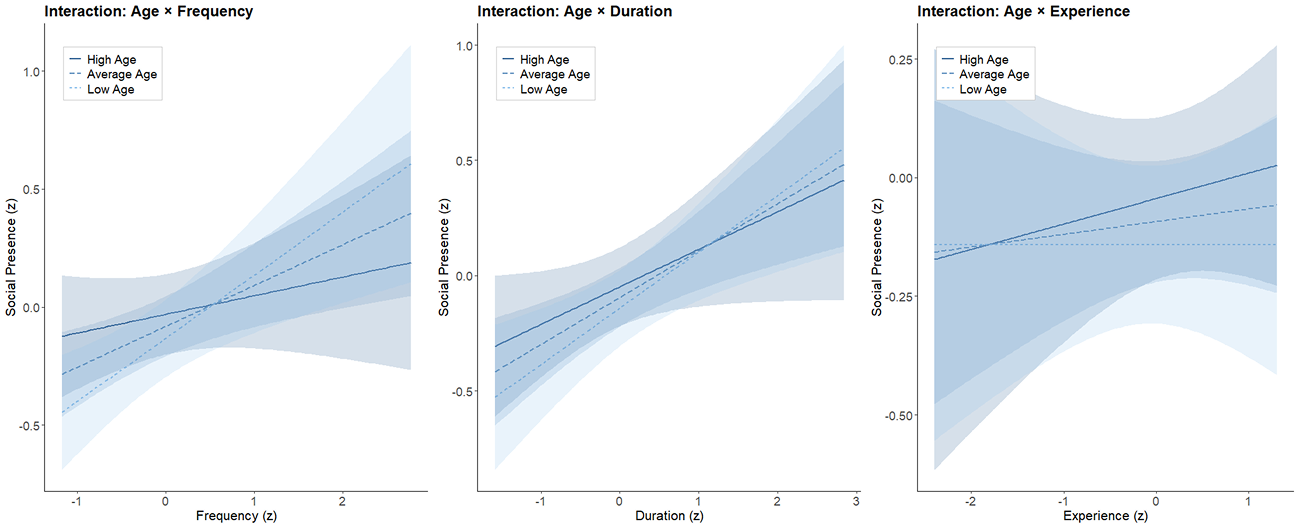}
    \caption{\textbf{Interaction effects between age and usage behaviors on sense of presence.
    (Left) Age $\times$ Frequency, (Middle) Age $\times$ Duration, (Right) Age $\times$ Experience. 
    Shaded regions represent 95\% confidence intervals.}}
    \label{fig:agesocial}
\end{figure}

Interaction terms involving gender and age were added in separate models to test for moderation, as shown in \autoref{tab:socialpresence-interaction}. None of the gender- or age-based interactions reached conventional levels of significance. Overall, these results indicate that the observed effects of usage behaviors on social presence are relatively stable across demographic groups. Visualizations of the gender- and age-based moderation models are available in \autoref{fig:gendersocial} and \autoref{fig:agesocial} (Appendix~\ref{AppendixD})

\subsection{Models Predicting Spatial Presence}

Following the same modeling structure, we analyzed predictors of spatial presence in social VR.

\begin{table*}[htbp]
\centering
\renewcommand{\arraystretch}{1.2}
\caption{\textbf{Hierarchical Regression Predicting Spatial Presence}}
\label{tab:spatial_presence_regression}

\resizebox{.7\textwidth}{!}{%
\begin{tabular}{lcccccccc}
\toprule
\textbf{Variable} & \multicolumn{2}{c}{\textbf{Model 1}} & \multicolumn{2}{c}{\textbf{Model 2}} & \multicolumn{2}{c}{\textbf{Model 3}} & \multicolumn{2}{c}{\textbf{Model 4}} \\
\cmidrule(lr){2-3} \cmidrule(lr){4-5} \cmidrule(lr){6-7} \cmidrule(lr){8-9}
 & $B$ & $\beta$ & $B$ & $\beta$ & $B$ & $\beta$ & $B$ & $\beta$ \\
\midrule
Female            & 0.328**  & 0.148 & 0.317**  & 0.143 & 0.322**  & 0.145 & 0.340*** & 0.154 \\
Age (z)           & 0.024    & 0.025 & 0.017    & 0.017 & 0.024    & 0.024 & 0.022    & 0.022 \\
Frequency (z)     & 0.171*** & 0.174 & 0.156*** & 0.159 & 0.168*** & 0.171 & 0.163*** & 0.166 \\
Duration (z)      & 0.158**  & 0.160 & 0.147**  & 0.149 & 0.153**  & 0.155 & 0.149**  & 0.152 \\
Years (z)         & 0.039    & 0.040 & 0.038    & 0.039 & 0.044    & 0.045 & 0.067    & 0.068 \\
Freq $\times$ Dur &          &       & 0.130**  & 0.141 &          &       &          &       \\
Freq $\times$ Yr  &          &       &          &       & 0.090*   & 0.093 &          &       \\
Dur $\times$ Yr   &          &       &          &       &          &       & 0.129**  & 0.136 \\
\midrule
$R^2$             & \multicolumn{2}{c}{0.096} & \multicolumn{2}{c}{0.116} & \multicolumn{2}{c}{0.105} & \multicolumn{2}{c}{0.114} \\
$\Delta R^2$      & \multicolumn{2}{c}{--}    & \multicolumn{2}{c}{0.020**} & \multicolumn{2}{c}{\qj{0.009$^{\dagger}$}} & \multicolumn{2}{c}{0.018**} \\
\bottomrule
\multicolumn{9}{l}{\footnotesize \textit{Note.}  $\dagger p<.10$, * $p < .05$, ** $p < .01$, *** $p < .001$. Standardized $\beta$ coefficients shown without significance markers.}
\end{tabular}}
\end{table*}

\subsubsection{Main and Interaction Effects of Usage Behaviors}

\autoref{tab:spatial_presence_regression} shows the results.
In Model 1, both frequency ($\beta = .174$, $p < .001$) and duration ($\beta = .160$, $p < .01$) were significant predictors of spatial presence. Female users again reported higher levels than males ($\beta = .148$, $p < .01$). Neither age nor years of VR use reached significance in this model.

Model 2 added the interaction between frequency and duration, which significantly improved model fit ($\beta = .141$, $p < .01$), suggesting that those who use VR frequently and for long sessions are more likely to experience higher spatial presence.

In Model 3, the interaction between frequency and years of VR use was also significant ($\beta = .093$, $p < .05$), indicating that frequent usage has a stronger effect among long-term users. Model 4 introduced the duration × years interaction, which again reached significance ($\beta = .136$, $p < .01$). These results consistently show that behavioral engagement has a cumulative effect, with seasoned users deriving more spatial presence from intensive use. The interaction plots are presented in \autoref{fig:usageinteraction7} in Appendix~\ref{AppendixD}.

\subsubsection{Moderation Effects of Gender and Age}

\begin{figure}[H]
    \centering
    \includegraphics[width=1\linewidth]{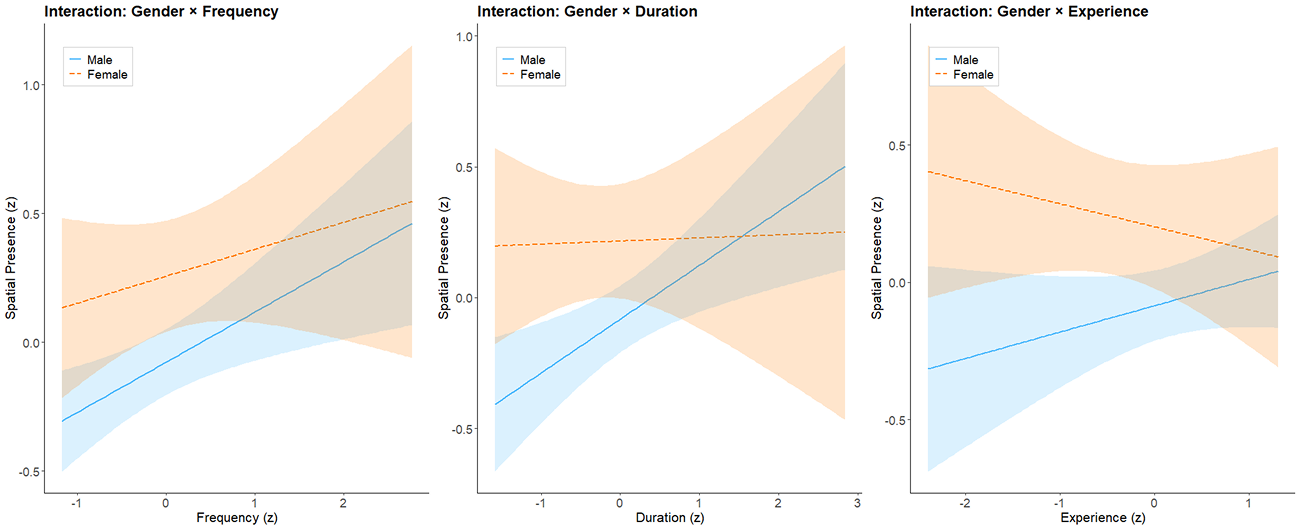}
    \caption{\textbf{Interaction effects between gender and usage behaviors on spatial presence.
    (Left) Gender $\times$ Frequency, (Middle) Gender $\times$ Duration, (Right) Gender $\times$ Experience. 
    Shaded regions represent 95\% confidence intervals.}}
    \label{fig:genderspatial}
\end{figure}

\begin{figure}[H]
    \centering
    \includegraphics[width=1\linewidth]{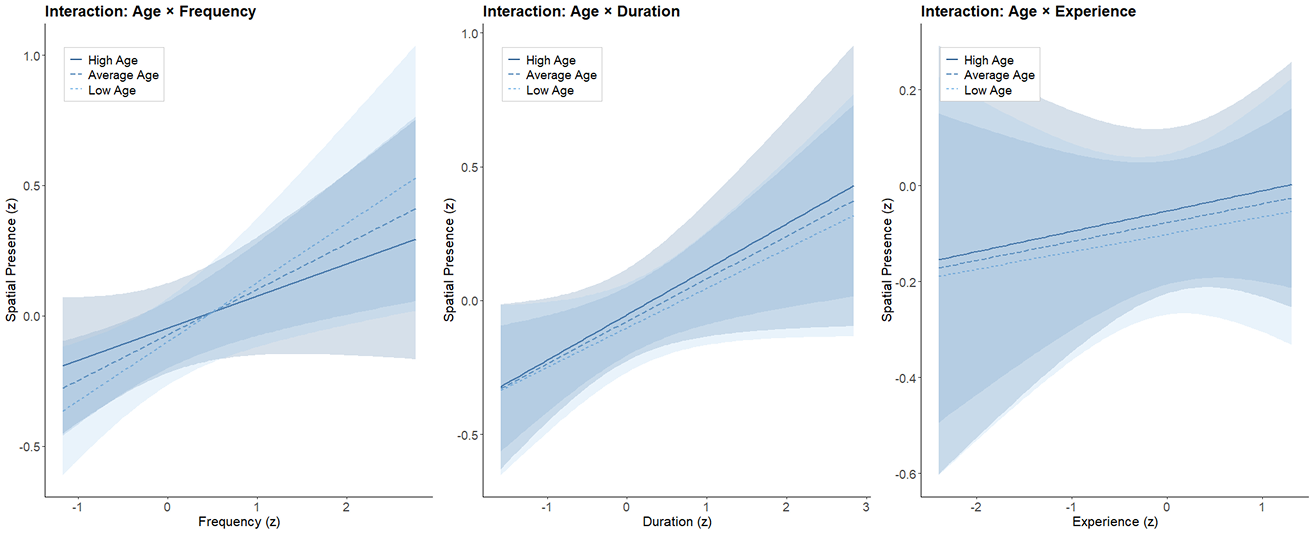}
    \caption{\textbf{Interaction effects between age and usage behaviors on spatial presence.
    (Left) Age $\times$ Frequency, (Middle) Age $\times$ Duration, (Right) Age $\times$ Experience. 
    Shaded regions represent 95\% confidence intervals.}}
    \label{fig:agespatial}
\end{figure}

\autoref{tab:spatialpresence-interaction} summarizes the moderation models involving gender and age.
None of the demographic interaction terms (e.g., Gender × Frequency, Age × Duration) were significant across the six moderation models. These findings suggest that the link between behavioral variables and spatial presence does not differ meaningfully across gender or age groups. Visualizations of moderation effects are provided in \autoref{fig:genderspatial} and \autoref{fig:agespatial} (Appendix~\ref{AppendixD}).

\begin{table*}[htbp]
\centering
\small
\setlength{\tabcolsep}{4pt}        
\renewcommand{\arraystretch}{1.1}  
\caption{\textbf{Hierarchical Regression: Gender and Age Interactions Predicting Spatial Presence}}
\label{tab:spatialpresence-interaction}
\begin{tabular}{lcccccccccccc}
\toprule
\textbf{Variable} & \multicolumn{2}{c}{\textbf{Model 1}} & \multicolumn{2}{c}{\textbf{Model 2}} & \multicolumn{2}{c}{\textbf{Model 3}} & \multicolumn{2}{c}{\textbf{Model 4}} & \multicolumn{2}{c}{\textbf{Model 5}} & \multicolumn{2}{c}{\textbf{Model 6}} \\
\cmidrule(lr){2-3} \cmidrule(lr){4-5} \cmidrule(lr){6-7} \cmidrule(lr){8-9} \cmidrule(lr){10-11} \cmidrule(lr){12-13}
 & $B$ & $\beta$ & $B$ & $\beta$ & $B$ & $\beta$ & $B$ & $\beta$ & $B$ & $\beta$ & $B$ & $\beta$ \\
\midrule
Female              & 0.334** & 0.151 & 0.298** & 0.135 & 0.287**  & 0.130 & 0.318** & 0.144 & 0.330** & 0.149 & 0.328** & 0.148 \\
Age (z)             & 0.021   & 0.021 & 0.020    & 0.020 & 0.021    & 0.022 & 0.026    & 0.027 & 0.024    & 0.025 & 0.024    & 0.025 \\
Frequency (z)       & 0.195***& 0.198 & 0.177*** & 0.179 & 0.176*** & 0.179 & 0.175*** & 0.178 & 0.170*** & 0.173 & 0.171*** & 0.174 \\
Duration (z)        & 0.159***& 0.161 & 0.206** & 0.209 & 0.152**  & 0.154 & 0.160*** & 0.162 & 0.159*** & 0.161 & 0.158**  & 0.160 \\
Years (z)           & 0.042   & 0.042 & 0.037    & 0.038 & 0.096    & 0.097 & 0.037    & 0.037 & 0.040    & 0.040 & 0.039    & 0.039 \\
G $\times$ Freq     & $-$0.090 & $-$0.048 &         &        &         &        &         &        &         &        &         &        \\
G $\times$ Dur      &          &         & $-$0.194 & $-$0.101 &         &        &         &        &         &        &         &        \\
G $\times$ Years    &          &         &          &         & $-$0.180 & $-$0.103 &         &        &         &        &         &        \\
A $\times$ Freq     &          &         &          &         &          &         & $-$0.052 & $-$0.051 &         &        &         &        \\
A $\times$ Dur      &          &         &          &         &          &         &          &         & 0.011    & 0.011 &         &        \\
A $\times$ Years    &          &         &          &         &          &         &          &         &          &         & 0.003    & 0.003 \\
\midrule
$R^2$               & \multicolumn{2}{c}{0.098} & \multicolumn{2}{c}{0.103} & \multicolumn{2}{c}{0.103} & \multicolumn{2}{c}{0.099} & \multicolumn{2}{c}{0.096} & \multicolumn{2}{c}{0.096} \\
$\Delta R^2$        & \multicolumn{2}{c}{0.002} & \multicolumn{2}{c}{0.005} & \multicolumn{2}{c}{0.005} & \multicolumn{2}{c}{0.003} & \multicolumn{2}{c}{0.000} & \multicolumn{2}{c}{0.000} \\
\bottomrule
\multicolumn{13}{l}{\footnotesize \textit{Note.} \qj{$\dagger p<.10$}, * $p < .05$, ** $p < .01$, *** $p < .001$. Standardized $\beta$ shown without significance markers.}
\end{tabular}
\end{table*}

\subsection{Models Predicting Self Presence}

Finally, we examined predictors of self presence—users’ sense of embodying or identifying with their avatar or virtual self.

\begin{table*}[t]
\centering
\renewcommand{\arraystretch}{1.2}
\caption{\textbf{Hierarchical Regression Predicting Self Presence}}
\label{tab:self_presence_regression}

\resizebox{.7\textwidth}{!}{%
\begin{tabular}{lcccccccc}
\toprule
\textbf{Variable} & \multicolumn{2}{c}{\textbf{Model 1}} & \multicolumn{2}{c}{\textbf{Model 2}} & \multicolumn{2}{c}{\textbf{Model 3}} & \multicolumn{2}{c}{\textbf{Model 4}} \\
\cmidrule(lr){2-3} \cmidrule(lr){4-5} \cmidrule(lr){6-7} \cmidrule(lr){8-9}
 & $B$ & $\beta$ & $B$ & $\beta$ & $B$ & $\beta$ & $B$ & $\beta$ \\
\midrule
Female            & 0.209       & 0.094  & 0.195       & 0.087  & 0.201       & 0.090  & 0.222$^{\dagger}$ & 0.099 \\
Age (z)           & 0.072       & 0.073  & 0.062       & 0.063  & 0.072       & 0.072  & 0.070       & 0.070 \\
Frequency (z)     & 0.191***    & 0.192  & 0.171***    & 0.173  & 0.186***    & 0.188  & 0.182***    & 0.184 \\
Duration (z)      & 0.131**     & 0.131  & 0.116*      & 0.117  & 0.125**     & 0.125  & 0.122**     & 0.123 \\
Years (z)         & $-0.034$    & $-0.034$ & $-0.035$  & $-0.035$ & $-0.028$  & $-0.028$ & $-0.004$  & $-0.004$ \\
Freq $\times$ Dur &             &        & 0.171***    & 0.183  &             &        &             &       \\
Freq $\times$ Yr  &             &        &             &        & 0.119**     & 0.122  &             &       \\
Dur $\times$ Yr   &             &        &             &        &             &        & 0.136**     & 0.141 \\
\midrule
$R^2$             & \multicolumn{2}{c}{0.080} & \multicolumn{2}{c}{0.112} & \multicolumn{2}{c}{0.094} & \multicolumn{2}{c}{0.099} \\
$\Delta R^2$      & \multicolumn{2}{c}{--}    & \multicolumn{2}{c}{0.032**} & \multicolumn{2}{c}{0.014*} & \multicolumn{2}{c}{0.019*} \\
\bottomrule
\multicolumn{9}{l}{\footnotesize \textit{Note.} $\dagger p<.10$, * $p<.05$, ** $p<.01$, *** $p<.001$. Standardized $\beta$ coefficients shown without significance markers.}
\end{tabular}}
\end{table*}

\subsubsection{Main and Interaction Effects of Usage Behaviors}
\autoref{tab:self_presence_regression} summarizes the results.
Model 1 showed that frequency ($\beta = .192$, $p < .001$) and duration ($\beta = .131$, $p < .01$) were both positively associated with self presence. 
Model 2 added the interaction between frequency and duration, which significantly predicted self presence ($\beta = .183$, $p < .001$). This finding highlights a compounding effect: engaging frequently and for longer sessions enhances one’s embodied experience in VR.

\begin{figure}[H]
    \centering
    \includegraphics[width=1\linewidth]{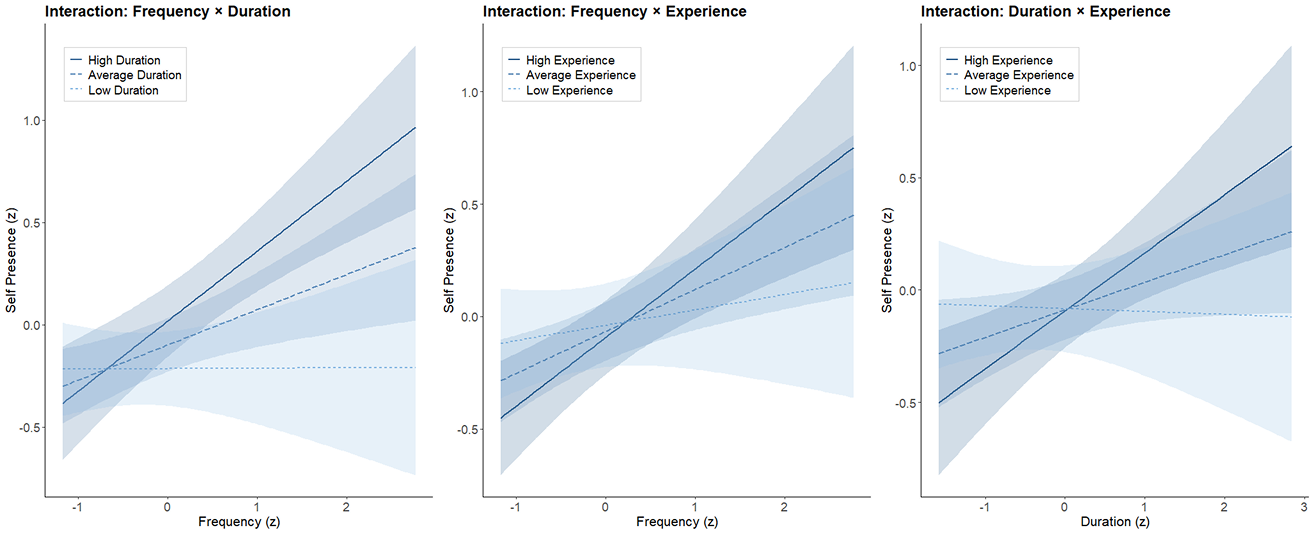}
     \caption{\textbf{Interaction effects of usage behaviors on self presence. 
    (Left) Frequency $\times$ Duration interaction; (Middle) Frequency $\times$ Experience interaction; (Right) Duration $\times$ Experience interaction. 
    All predictors are standardized. Shaded areas represent 95\% confidence intervals.}}
    \label{fig:usageinteraction}
\end{figure}

Model 3 introduced the frequency × years interaction, also significant ($\beta = .122$, $p < .01$), indicating that frequency of use is especially impactful among experienced users. Model 4 tested the duration × years interaction and again found a significant effect ($\beta = .141$, $p < .01$). These results collectively point to a pattern where long-term, intensive use of VR technologies strengthens users’ identification with their virtual selves.
Further illustrations of these interaction models can be found in \autoref{fig:usageinteraction} (Appendix~\ref{AppendixD}).

\begin{table*}[htbp]
\centering
\small
\setlength{\tabcolsep}{4pt}        
\renewcommand{\arraystretch}{1.1} 
\caption{\textbf{Hierarchical Regression: Gender and Age Interactions Predicting Self Presence}}
\label{tab:selfpresence-interaction}
\begin{tabular}{lcccccccccccc}
\toprule
\textbf{Variable} & \multicolumn{2}{c}{\textbf{Model 1}} & \multicolumn{2}{c}{\textbf{Model 2}} & \multicolumn{2}{c}{\textbf{Model 3}} & \multicolumn{2}{c}{\textbf{Model 4}} & \multicolumn{2}{c}{\textbf{Model 5}} & \multicolumn{2}{c}{\textbf{Model 6}} \\
\cmidrule(lr){2-3} \cmidrule(lr){4-5} \cmidrule(lr){6-7} \cmidrule(lr){8-9} \cmidrule(lr){10-11} \cmidrule(lr){12-13}
 & $B$ & $\beta$ & $B$ & $\beta$ & $B$ & $\beta$ & $B$ & $\beta$ & $B$ & $\beta$ & $B$ & $\beta$ \\
\midrule
Female              & 0.218$^\dagger$ & 0.099 & 0.183    & 0.082 & 0.182    & 0.081 & 0.198    & 0.088 & 0.204    & 0.094 & 0.210    & 0.094 \\
Age (z)             & 0.068           & 0.068 & 0.068    & 0.071 & 0.070    & 0.073 & 0.075    & 0.075 & 0.072    & 0.072 & 0.073    & 0.073 \\
Frequency (z)       & 0.225***        & 0.226 & 0.195*** & 0.195 & 0.194*** & 0.195 & 0.195*** & 0.196 & 0.194*** & 0.195 & 0.191*** & 0.191 \\
Duration (z)        & 0.132**         & 0.133 & 0.172**  & 0.173 & 0.127**  & 0.128 & 0.133**  & 0.133 & 0.128**  & 0.128 & 0.131**  & 0.131 \\
Years (z)           & $-$0.036        & $-$0.036 & 0.004    & 0.004 & $-$0.037 & $-$0.037 & $-$0.034 & $-$0.034 & $-$0.034 & $-$0.034 & $-$0.034 & $-$0.034 \\
G $\times$ Freq     & $-$0.127        & $-$0.067 &         &        &         &        &         &        &         &        &         &        \\
G $\times$ Dur      &                &        & $-$0.166 & $-$0.086 &         &        &         &        &         &        &         &        \\
G $\times$ Years    &                &        &         &        & $-$0.119 & $-$0.068 &         &        &         &        &         &        \\
A $\times$ Freq     &                &        &         &        &         &        & $-$0.061 & $-$0.058 &         &        &         &        \\
A $\times$ Dur      &                &        &         &        &         &        &         &        & $-$0.032 & $-$0.032 &         &        \\
A $\times$ Years    &                &        &         &        &         &        &         &        &         &        & 0.009    & 0.009 \\
\midrule
$R^2$               & \multicolumn{2}{c}{0.083} & \multicolumn{2}{c}{0.085} & \multicolumn{2}{c}{0.083} & \multicolumn{2}{c}{0.083} & \multicolumn{2}{c}{0.081} & \multicolumn{2}{c}{0.080} \\
$\Delta R^2$        & \multicolumn{2}{c}{0.003} & \multicolumn{2}{c}{0.005} & \multicolumn{2}{c}{0.003} & \multicolumn{2}{c}{0.003} & \multicolumn{2}{c}{0.001} & \multicolumn{2}{c}{0.000} \\
\bottomrule
\multicolumn{13}{l}{\footnotesize \textit{Note.} $\dagger p<.10$, * $p < .05$, ** $p < .01$, *** $p < .001$. Standardized $\beta$ shown without significance markers.}
\end{tabular}
\end{table*}

\subsubsection{Moderation Effects of Gender and Age}
As with previous dimensions, we tested whether the effects of usage behaviors on self presence varied across gender or age. None of the interaction terms reached statistical significance, as visualized in \autoref{tab:selfpresence-interaction}. 

Overall, these results suggest that the relationship between behavioral usage and self presence is consistent across demographic subgroups. While usage frequency, session duration, and their combinations are important predictors, these effects do not meaningfully differ by gender or age.
Supplementary figures of these moderation effects are shown in \autoref{fig:genderself} and \autoref{fig:ageself} in Appendix~\ref{AppendixD}.

\section{Discussion}
In this section, we reflect on the main findings, starting with how usage intensity shapes the experience of presence. We then discuss the role of gender and age, and outline limitations alongside directions for future research.

\subsection{Role of Usage Intensity in Presence Formation}

Our findings demonstrate that usage intensity, specifically the frequency of use and the average duration per session, is the most robust predictor of users’ sense of presence in social VR environments. Across all four presence dimensions (overall, social, spatial, and self presence), both behavioral indicators showed consistently strong and significant associations with presence levels, while the length of VR experience exerted only weak or non-significant effects. 
 This pattern suggests that current behavioral engagement (how often and how long users actively participate in VR) plays a more influential role in shaping immersive experiences than the length of past exposure.
In other words, usage patterns, rather than users’ historical familiarity with VR, have a more powerful impact on their subjective sense of presence.

Beyond main effects, our analyses also uncovered a series of significant interaction effects between usage variables, highlighting the cumulative or 'amplifying' nature of usage intensity on presence. 
Interactions such as frequency × duration, frequency × years of use, and duration × years of use were statistically significant across multiple presence dimensions, particularly in the models for overall, spatial presence, and self presence. These results point to a synergistic pattern: when high-frequency use, extended session duration, and long-term experience co-occur (i.g, when any two forms or all of intensive usage co-occur), the resulting level of presence far exceeds the additive effect of any single factor leading to nonlinear gains. In practical terms, for example, users who engage in VR both frequently and for long durations report a substantially stronger sense of presence than would be predicted by each usage factor alone.
These findings provide empirical evidence supporting theoretical accounts that frame presence as an adaptive or cumulative phenomenon \citep{slater2003note, witmer1998measuring}. \citet{witmer1998measuring}, for instance, argue that multiple factors, such as focused attention, user involvement, and sensory fidelity, may act in concert to reinforce the illusion of non-mediation. Our results align with this perspective and extend it by showing that patterns of VR use also contribute synergistically, with usage intensity exerting non-linear effects on presence.

These findings inform VR system design and user engagement strategies, particularly for applications that benefit from high presence (such as educational simulations, therapy, or collaborative virtual worlds). First, the above effects suggest that designs encouraging frequent and longer usage sessions can significantly amplify presence. VR designers might consider features that promote regular engagement, such as daily login incentives, evolving content, or social features that entice users to return often. By increasing usage frequency, developers could help users build a habit of entering the virtual world, lowering the entry barrier to feeling present. Our results hint, the frequency should be better paired with quality immersive experiences of substantial length. Therefore, VR applications should also strive to support longer session durations without causing fatigue. This might involve improving user comfort (e.g. minimizing motion sickness through technical measures) and providing options for users to take breaks or manage their level of stimulation. For instance, since repeated exposure can reduce cybersickness over time, onboarding programs could gradually increase session length as users acclimate. Designers should craft experiences that become more engaging over time, rewarding extended presence, such as narrative-driven environments or open-ended worlds that unfold with longer play.

\subsection{Demographic Differences in Presence}
Our findings reveal both meaningful differences and notable consistencies in how demographic factors (particularly gender and age) relate to users’ sense of presence in social VR environments.

\subsubsection{Gender Differences}
Notably, this gender gap emerged as an overall difference rather than a divergent response to usage behaviors, since none of the gender × behavior interaction effects (e.g. gender × frequency of use) were significant. In other words, women and men appear to respond similarly to how VR usage intensity or activities affect presence. 

But the findings indicate that female users reported a higher presence overall, with significant differences in the social and spatial dimensions, and a marginal tendency toward higher self-presence.
This contrasts with much of the prior work on gender and presence, which has largely focused on spatial presence and sometimes found men reporting stronger immersion \citep{yang2024exploring}. Our results suggest that, in naturalistic multi-user settings, presence is multidimensional and gendered differences can extend beyond social engagement to include spatial immersion as well.

One possible explanation is that women engage more strongly with the social aspect of VR. Prior work shows that women often adopt a cooperative, connection-oriented communication style \citep{menshikova2018gender, royal1972non, toussaint2005gender}, and in virtual environments they tend to maintain closer distances between avatars, indicating greater social engagement. Such tendencies may amplify social presence.

Differences with earlier research may stem from the fact that earlier research was typically conducted in laboratory settings and often treated presence as a single, unified construct. In such controlled environments, social interactions tend not to occur naturally. Moreover, identity management and exploration (both crucial for fostering self-presence) are rarely supported.
 By contrast, multi-user VR environments prioritize natural social interactions, identity play, and the exchange of rich social cues \citep{freeman2021body, maloney2020falling, Hugging5}. Within this more ecological context, women’s strengths in social orientation may also translate into stronger spatial immersion.

 These results contribute to ongoing discussions about the role of gender in shaping VR experiences. They suggest that gender differences in presence may be more about initial dispositions or social orientation. We encourage future research to further explore why female users report higher social and spatial presence, possibly examining gender-linked factors like empathy or social cue sensitivity, and how virtual environments might be tailored to accommodate these differences without reinforcing stereotypes.

\subsubsection{Age Differences} 

The present study found that age was not a significant predictor of presence in a multi-user social VR setting, nor did age interact with user behaviors to influence presence. This null finding must be contextualized against a background of mixed results in the literature. On one hand, some prior studies suggested that presence tends to decline with increasing age, implying that older adults feel less “immersed” or “being there” in VR than younger users \cite{kober2014effects}. On the other hand, a growing number of studies have reported little to no age effect on presence, or even the reverse trend \cite{dilanchian2021pilot, Lorenz2023Age}. This divergence in findings suggests that the relationship between age and presence is more complex than a simple linear decline. Our results align more closely with studies showing minimal or no age differences, reinforcing the idea that being older does not inherently diminish one’s capacity to experience presence in virtual environments. 
One likely explanation for the conflicting evidence is that age itself is not a direct determinant of presence, but rather an indirect factor operating through various mediators. 

A mediator is that environmental familiarity may mediate the influence of age on presence. Younger users often have more extensive gaming or tech experience, which could facilitate quick acclimation to VR controls and conventions; older users might initially face usability challenges, potentially dampening presence until they overcome those barriers \cite{harrington2017developing}. This lack of familiarity many initially manifest as disorientation or a need to consciously think about the interface, which can detract from the immediacy of the VR experience. Kober’s study that found an age-related decline in presence has been critiqued on these grounds: the older group
likely had far less VR exposure than the younger \citep{mitzner2021understanding}. In our study, with a presumably self-selected sample of users who engage in social VR as opposite with laboratory settings, older participants may have already surmounted many of these initial hurdles, resulting in no observable presence gap. This points to a theoretical interpretation: age influences how people approach VR (their learning curve, comfort with technology, genre preferences), rather than directly changing their ability to experience presence. Differences labeled as “age effects” in some studies may have been capturing differences in technology literacy or learning curve rather than age per se.

An additional mediator to consider is interest and motivation, which ties into content relevance. Presence is not only a product of technology and cognition, but also of how much the user wants to be in the virtual world. Differences in cultural context and personal interests across age groups can therefore modulate presence. Some earlier studies implicitly used content geared toward younger audiences (e.g. fast-paced video games or abstract tasks), which might fail to captivate older participants. In fact, attitudinal barriers have been documented: some older adults perceive video games and VR as frivolous or “not for people like me,” and they may worry about being seen as childish for participating \citep{de2015domestication}. Such attitudes can dampen the enthusiasm and openness that are critical for presence. By contrast, older adults report greater engagement when applications are perceived as valuable. For example, when they support social connection, education, or health and well-being \citep{de2015domestication}.
Our study’s multi-user social context likely provided exactly this kind of meaningful engagement. 
The naturalistic scenario may have allowed older participants to draw on familiar social skills and focus on the interpersonal experience instead of the technology, thereby enhancing their sense of being really there. In summary, theoretical reasoning and evidence converge on a key point: age influences presence only through mediated pathways (e.g., cognitive load and interest), and when those factors are accounted for or optimized, the simple effect of age may disappear. This perspective helps explain why our findings align with some studies and contradict others – those differences likely hinged on how demanding the VR experience was, how comfortable the participants were with the medium, and how invested they were in the content.

These findings speak to both research and design. From a research perspective, they highlight the importance of examining the mechanisms through which age influences presence, rather than treating age as a fixed determinant. Presence models could benefit from incorporating factors such as expertise, interest, and cognitive strategies: dimensions that may change with age but are not immutable. It may be fruitful for future work to examine, for example, how social facilitation in VR (having other people present) might differentially affect users who are less technologically adept, or how training interventions could reduce the age gaps in presence. Practically, the results suggest that VR can serve as an inclusive medium across the lifespan. In a context of global population aging, it is encouraging that older participants in our study reported presence levels comparable to younger peers. Rather than excluding older adults from VR experiences or designing separate, watered-down applications for them, efforts may be better spent ensuring that mainstream VR platforms are both accessible and engaging for diverse age groups. Older adults will engage deeply with VR when the context is right.

\subsection{Limitations and Future Works}
Although this study provides evidence on how usage intensity shapes different dimensions of presence in social VR, certain limitations should be noted. The sample was relatively young and primarily based in Europe, which may affect the extent to which the findings generalize to broader populations. 
\qj{In addition, our sample included more men than women. While such an imbalance does not bias the regression coefficient for gender, it reduces the precision of estimates for the smaller group. The main effect for women should therefore be interpreted with some caution.}

\qj{Besides, 
our sample more likely represents users who voluntarily engage with social VR and are sufficiently comfortable with VR to continue using it, while underrepresenting those who use VR in more constrained or mandatory settings, or who disengage because of discomfort. Consequently, the present results are most directly applicable to voluntary, everyday social VR use, and their applicability to professional, medical, or educational deployments of VR should be interpreted with some caution. 
} 

\qj{A further limitation concerns how presence was measured. Our survey items capture participants’ general experience in social VR, and thus assess recalled typical levels of spatial, social, and self-presence across multiple sessions rather than immediate, within-session states. This retrospective, generalized approach is common in large-scale survey work and has been utilized to measure presence in both social VR and other online environments \cite{van2023feelings,barreda2022hooked, MirroredPresence}. However, such reports can be shaped by memory and reconstruction processes (e.g., decay or how recently participants last engaged with VR).} 
\qj{An additional related limitation is that the measures do not allow us to draw conclusions about within-session fluctuations in presence during specific VR encounters. The findings should therefore be interpreted as relationships between usage intensity and users’ typical sense of presence in social VR. They do not speak directly to within-session, state-like experiences during individual VR encounters.}

Building on these findings, directions emerge for future research. First, while our models captured usage patterns, future studies could incorporate more behavioral metrics (e.g., quantity of social interactions). 
Second, although usage patterns predicted presence, the mechanisms linking usage to immersive experiences remain unclear. Future studies could examine mediating factors, such as: Flow or deep task engagement; Social connection and empathy; Cognitive absorption or attention allocation.
Such models would help unpack why frequent or long use translates into higher presence.
Additionally, the consistent interaction effects suggest that presence may be non-linearly enhanced by combining behavioral factors (e.g., frequency and duration). Future systems could dynamically adapt to users’ patterns, offering personalized recommendations for usage pacing, break scheduling, or content delivery that optimize presence while avoiding fatigue or overload.
\qj{Our measures of usage intensity are based on self-report questionnaires rather than logged behavior. 
Future work should combine behavioral logs and self-reports to obtain a more robust, multimethod assessment of presence.}
\qj{Finally, validating these patterns in samples with more balanced gender representation and more diverse age and geographic backgrounds would help strengthen the generalizability of the findings.}


\section{Conclusion}
This study investigated how usage intensity, captured through frequency, session duration, and years of experience, shapes users’ sense of presence in social VR environments. The findings demonstrate that both frequency and duration of use are significant positive predictors of presence, underscoring the importance of consistent and prolonged engagement in fostering embodied experience. Moreover, the interaction between frequency and duration suggests a compounding effect: users who engage both frequently and for longer sessions experience higher levels of presence than would be expected from either factor alone.
Although gender showed a main effect, with female users reporting higher presence, no significant interaction effects were found for gender or age, suggesting that these relationships hold consistently across demographic subgroups in our sample. 
By highlighting the behavioral antecedents of sense of presence, this study contributes to a deeper understanding of how immersive experiences unfold in everyday social VR use beyond tightly controlled lab environments. The findings point to design opportunities for virtual environments that adapt to users’ engagement patterns and support inclusive experiences. 



\begin{acks} 
We thank the anonymous reviewers for their constructive feedback, which helped us significantly improve the paper. This work was funded by the Research Council of Finland (Grant \#357270).
\end{acks}

\bibliographystyle{ACM-Reference-Format}
\bibliography{sample-base}


\appendix

\section{Appendix}
\label{AppendixA}
\section*{Participant Demographics.}

Table~\ref{tab:demographics} presents the demographic profile of the study participants ($N = 295$). 

\begin{table}[h]
\centering
\renewcommand{\arraystretch}{1.2}
\caption{\textbf{Participant Demographics: Gender, Age, and Geographic Distribution (N = 295)}}
\label{tab:demographics}
\begin{tabular}{llcc}
\toprule
\textbf{Category} & \textbf{Group} & \textbf{Count} & \textbf{\%} \\
\midrule
\multirow{4}{*}{Gender} 
    & Man               & 215 & 72.9 \\
    & Woman             & 79  & 26.8 \\
    & Non-binary         & 1   & 0.3  \\
    & Prefer not to say  & 0   & 0.0  \\
\midrule
\multirow{7}{*}{Age} 
    & 18 or under        & 0   & 0.0  \\
    & 18--24 years       & 54  & 18.3 \\
    & 25--34 years       & 140 & 47.5 \\
    & 35--44 years       & 61  & 20.7 \\
    & 45--54 years       & 31  & 10.5 \\
    & 55--64 years       & 6   & 2.0  \\
    & 65 or above        & 3   & 1.0  \\
\midrule
\multirow{5}{*}{Continent} 
    & Africa             & 29  & 9.8  \\
    & Americas           & 53  & 18.0 \\
    & Asia               & 8   & 2.7  \\
    & Europe             & 200 & 67.8 \\
    & Oceania            & 5   & 1.7  \\
\bottomrule
\end{tabular}
\end{table}

\section{Appendix}
\label{AppendixB}

\section*{VR Usage Patterns}

Table~\ref{tab:vr_usage_patterns} provides a descriptive overview of participants' VR usage behaviors, including frequency of use, average session duration, and total experience with social VR systems.

\begin{table}[h]
\centering
\caption{\textbf{VR Usage Patterns: Frequency, Duration, and Years of Use}}
\label{tab:vr_usage_patterns}
\renewcommand{\arraystretch}{1.2}
\small
\begin{tabular}{llcc}
\toprule
\textbf{Category} & \textbf{Group} & \textbf{Count} & \textbf{\%} \\
\midrule
\multirow{5}{*}{Frequency} 
& Not at all              & 0   & 0.0  \\
& Once a month or less    & 100 & 33.9 \\
& A few times a month     & 130 & 44.1 \\
& A few times a week      & 62  & 21.0 \\
& Daily                   & 3   & 1.0  \\
\midrule
\multirow{5}{*}{Session Duration} 
& < 30 minutes            & 37  & 12.5 \\
& 30--60 minutes          & 134 & 45.4 \\
& 1--2 hours              & 89  & 30.2 \\
& 2--3 hours              & 29  & 9.8  \\
& > 3 hours               & 6   & 2.0  \\
\midrule
\multirow{5}{*}{Years of VR Use} 
& < 1 month               & 9   & 3.1  \\
& 1--6 months             & 46  & 15.6 \\
& 6 months to 1 year      & 65  & 22.0 \\
& 1--3 years              & 112 & 38.0 \\
& > 3 years               & 63  & 21.4 \\
\bottomrule
\end{tabular}
\end{table}

\section{Appendix}
\label{AppendixC}
\section*{Measurement Properties of Presence Constructs}

This appendix summarizes the psychometric properties of the presence subscales used in the current study, including standardized factor loadings, internal consistency (Cronbach’s alpha), composite reliability (CR), and average variance extracted (AVE). All constructs demonstrated acceptable reliability and convergent validity.

\begin{table*}[h]
\centering
\renewcommand{\arraystretch}\small
\setlength{\tabcolsep}{4pt}        
\renewcommand{\arraystretch}{1.1}  
\caption{\textbf{Measurement Properties of Presence Constructs}}
\label{tab:presence-measurements}
\begin{tabular}{llccc}
\toprule
\textbf{Construct} & \textbf{Item} & \textbf{Label} & \textbf{Loading} & \textbf{Alpha / CR / AVE} \\
\midrule
\multirow{5}{*}{Social Presence (SP)} 
& SP1: I feel I am in the presence of other people. & SP1 & 0.737 & \multirow{5}{*}{0.855 / 0.857 / 0.548} \\
& SP2: Others in the environment are aware of my presence. & SP2 & 0.715 & \\
& SP3: The interface disappears; I interact with people. & SP3 & 0.720 & \\
& SP4: I interact with people, not simulations. & SP4 & 0.771 & \\
& SP5: Others appear conscious and alive. & SP5 & 0.756 & \\
\midrule
\multirow{4}{*}{Spatial Presence (SPP)} 
& SPP1: I feel I am actually in the VR environment. & SPP1 & 0.767 & \multirow{4}{*}{0.873 / 0.880 / 0.652} \\
& SPP2: I feel I take part in what is happening. & SPP2 & 0.739 & \\
& SPP3: My true location has shifted into VR. & SPP3 & 0.821 & \\
& SPP4: I feel physically present in the environment. & SPP4 & 0.861 & \\
\midrule
\multirow{5}{*}{Self Presence (SEL)} 
& SEL1: My avatar is an extension of my body. & SEL1 & 0.714 & \multirow{5}{*}{0.908 / 0.911 / 0.675} \\
& SEL2: What happens to avatar feels like my body. & SEL2 & 0.772 & \\
& SEL3: My real arm is projected into the avatar. & SEL3 & 0.841 & \\
& SEL4: My real hand is inside the virtual space. & SEL4 & 0.882 & \\
& SEL5: My avatar and body become one. & SEL5 & 0.876 & \\
\bottomrule
\end{tabular}
\end{table*}

\section{Appendix}
\label{AppendixD}

\section*{Interaction Effects: Visualizations}

This appendix presents graphical illustrations of interaction effects identified in the hierarchical regression analyses across four dimensions of presence: overall presence, social presence, spatial presence, and self presence. Figures are organized by dimension and type of interaction (usage behavior × usage behavior, gender × behavior, age × behavior). Each figure contains simple slopes visualized at ±1 SD levels of the moderator variable.

\subsection*{D.1 Interaction Effects on Overall Presence}

\begin{figure}[H]
    \centering
    \includegraphics[width=1\linewidth]{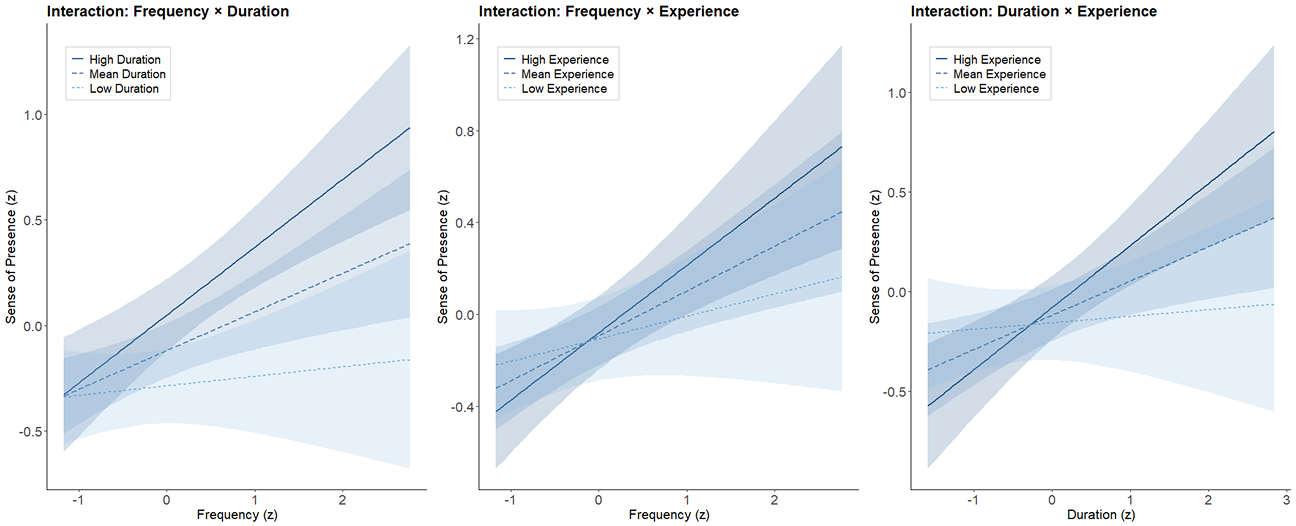}
    \caption{\textbf{Interaction effects of usage behaviors on sense of presence. 
    (Left) Frequency $\times$ Duration interaction; (Middle) Frequency $\times$ Experience interaction; (Right) Duration $\times$ Experience interaction. 
    All predictors are standardized. Shaded areas represent 95\% confidence intervals.}}
    \label{fig:interaction-effects}
\end{figure}

\subsection*{D.2 Interaction Effects on Social Presence}

\subsection*{D.3 Interaction Effects on Spatial Presence}

\begin{figure}[H]
    \centering
    \includegraphics[width=1\linewidth]{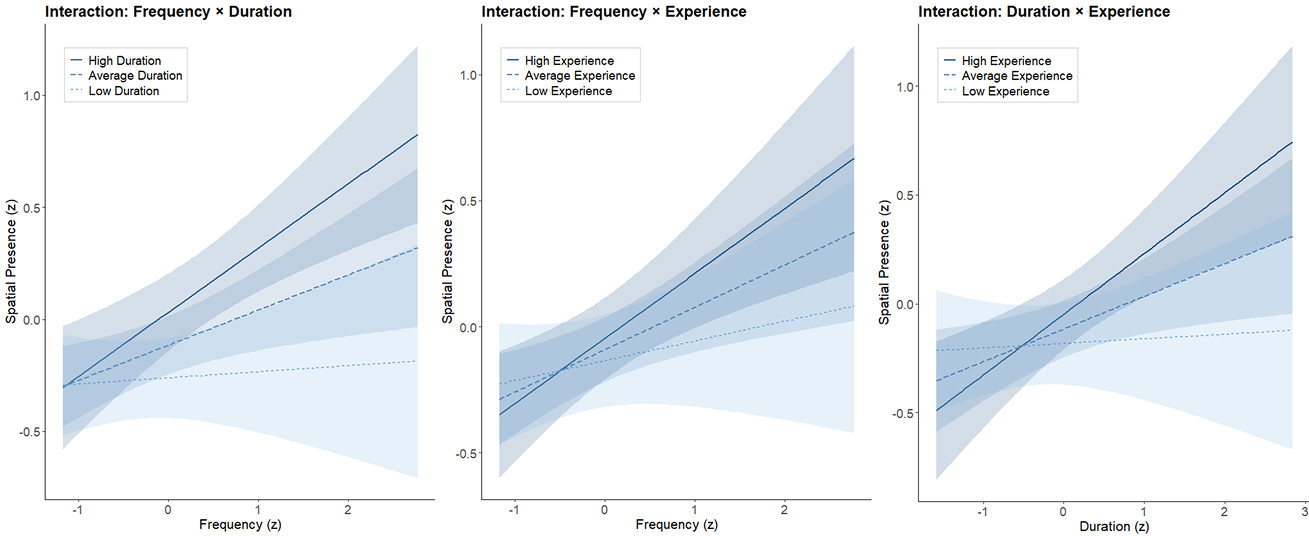}
    \caption{\textbf{Interaction effects of usage behaviors on spatial presence. 
    (Left) Frequency $\times$ Duration interaction; (Middle) Frequency $\times$ Experience interaction; (Right) Duration $\times$ Experience interaction. 
    All predictors are standardized. Shaded areas represent 95\% confidence intervals.}}
    \label{fig:usageinteraction7}
\end{figure}

\subsection*{D.4 Interaction Effects on Self Presence}

\begin{figure}[H]
    \centering
    \includegraphics[width=1\linewidth]{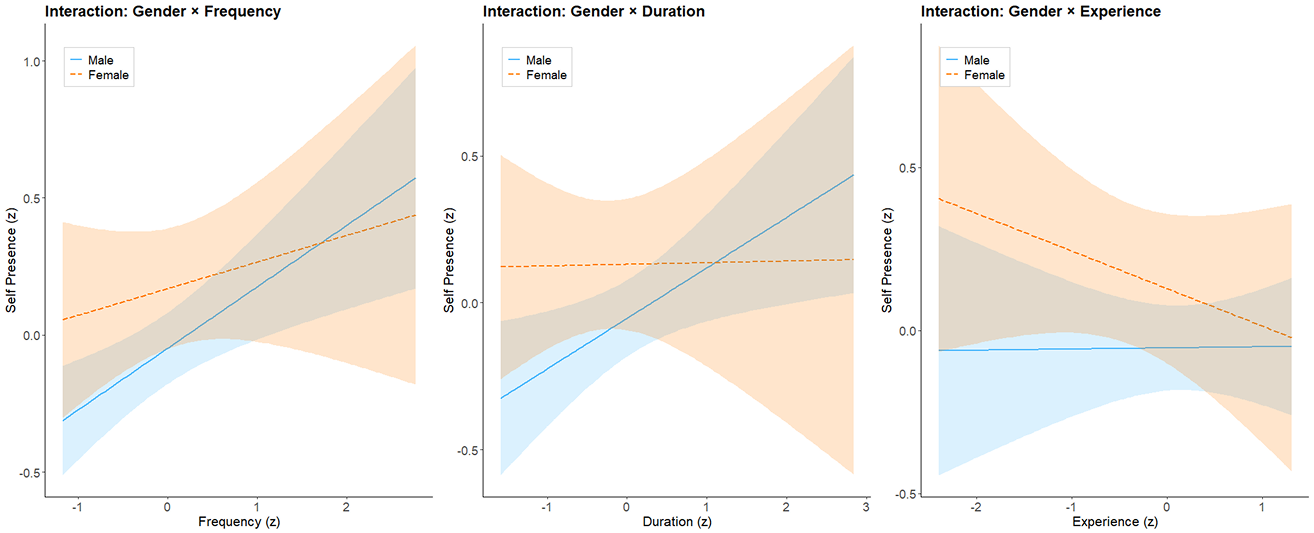}
    \caption{\textbf{Interaction effects between gender and usage behaviors on self presence.
    (Left) Gender $\times$ Frequency, (Middle) Gender $\times$ Duration, (Right) Gender $\times$ Experience. 
    Shaded regions represent 95\% confidence intervals.}}
    \label{fig:genderself}
\end{figure}

\begin{figure}[H]
    \centering
    \includegraphics[width=1\linewidth]{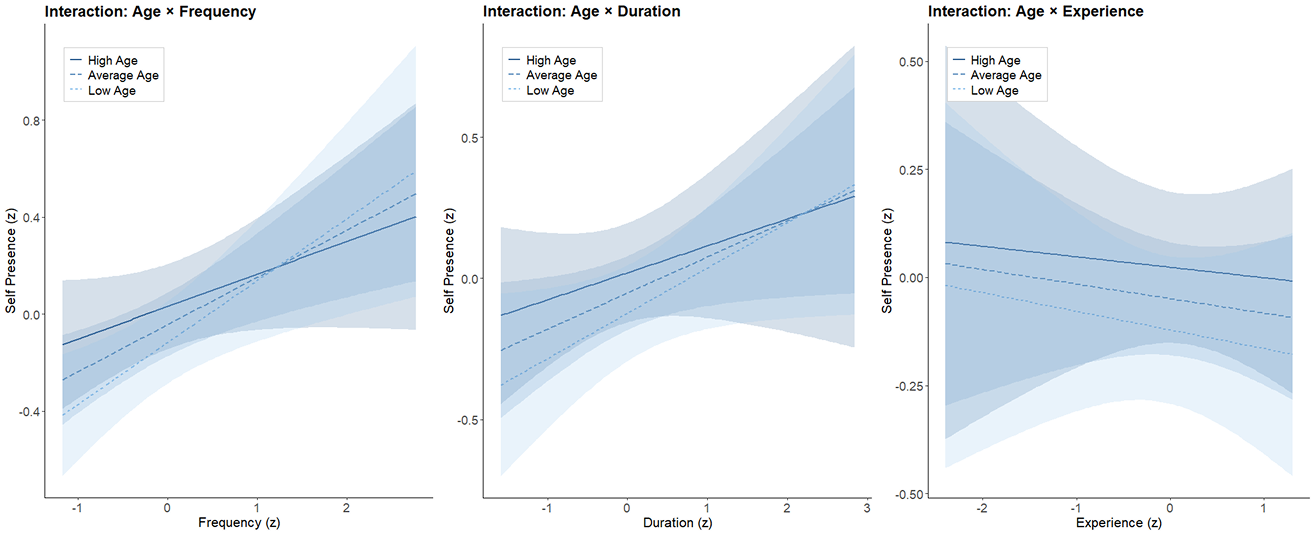}
    \caption{\textbf{Interaction effects between age and usage behaviors on self presence.
    (Left) Age $\times$ Frequency, (Middle) Age $\times$ Duration, (Right) Age $\times$ Experience. 
    Shaded regions represent 95\% confidence intervals}.}
    \label{fig:ageself}
\end{figure}

\end{document}